\documentclass[paper=a4, fontsize=11pt]{scrartcl} 

\usepackage[sort&compress,numbers]{natbib}
\usepackage{graphicx,epsfig}
\usepackage{algorithm,algorithmic}
\usepackage[mode=buildnew]{standalone}
\usepackage{tikz,pgfplots,subfig}
\pgfplotsset{compat=newest}
\usetikzlibrary{plotmarks}
\usetikzlibrary{positioning,arrows}
\usetikzlibrary{shapes.symbols,arrows,shadows,calc,positioning,pgfplots.groupplots,spy,backgrounds}
\usepackage{amsmath}
\usepackage{enumitem}

\usepackage[T1]{fontenc} 
\usepackage{fourier} 
\usepackage[english]{babel} 
\usepackage{amsmath,amsfonts,amsthm,amssymb} 
\usepackage{theorem}

\usepackage{lipsum} 

\usepackage{sectsty} 
\allsectionsfont{\centering \normalfont\scshape} 

\usepackage{fancyhdr} 
\numberwithin{equation}{section} 
\numberwithin{figure}{section} 
\numberwithin{table}{section} 

\newcommand{\horrule}[1]{\rule{\linewidth}{#1}} 

\title{
\normalfont \normalsize 
\textsc{ENSEEIHT} \\ [25pt] 
\horrule{0.5pt} \\[0.4cm] 
\huge EEG reconstruction and skull conductivity estimation using a Bayesian model promoting structured sparsity\\ 
\horrule{2pt} \\[0.5cm] 
}

\author{Facundo Costa, Hadj Batatia, Thomas Oberlin, Jean-Yves Tourneret} 

\date{\normalsize\today} 

\def\PM{\kern0pt^{\textrm{{\scriptsize PM}}}\kern0pt}
\def\MMAP{\kern1pt^{\textrm{{\tiny MMAP}}}\kern-1pt}

\def\MAP{^{\textrm{{\tiny MAP}}}}

  \def\cb{{\sbm{c}}\XS}
  
\def\Eb{{\sbm{E}}\XS}

\def\Hb{{\sbm{H}}\XS}  \def\hb{{\sbm{h}}\XS}
\def\Ib{{\sbm{I}}\XS}

\def\Mb{{\sbm{M}}\XS}  \def\mb{{\sbm{m}}\XS}

  \def\pb{{\sbm{p}}\XS}
  
  \def\rb{{\sbm{r}}\XS}
  \def\sb{{\sbm{s}}\XS}

  \def\vb{{\sbm{v}}\XS}
  
\def\Xb{{\sbm{X}}\XS}  \def\xb{{\sbm{x}}\XS}
\def\Yb{{\sbm{Y}}\XS}  \def\yb{{\sbm{y}}\XS}
  \def\zb{{\sbm{z}}\XS}
\def\mub{{\sbm{\mu}}\XS}

 \def\+{^\dagger}

\usepackage{afterpage}
\usepackage{textcomp}
\RequirePackage{xspace}

\newcommand{\diag}{\ensuremath{\mathrm{diag}}}
\newcommand{\II}{\ensuremath{\mathbb I}}

\newcommand{\RR}{\ensuremath{\mathbb R}}

\theoremstyle{plain}{\theorembodyfont{\rmfamily}%
\theoremstyle{plain}{\theorembodyfont{\rmfamily}%

\def\qed{\ifmmode\hbox{\hfill\sqb}\else{\ifhmode\unskip\fi%
\nobreak\hfil
\penalty50\hskip1em\null\nobreak\hfil$\blacksquare$
\parfillskip=0pt\finalhyphendemerits=0\endgraf}\fi}

\RequirePackage{amsmath}
\RequirePackage{xspace}
\RequirePackage{bbm}

\def\XS{\xspace}
\DeclareMathAlphabet{\mathb}{OML}{cmm}{b}{it}
\def\sbm#1{\ensuremath{\mathb{#1}}}                
\def\sbmm#1{\ensuremath{\boldsymbol{#1}}}          

  \def\cb{{\sbm{c}}\XS}
  
\def\Eb{{\sbm{E}}\XS}

\def\Hb{{\sbm{H}}\XS}  \def\hb{{\sbm{h}}\XS}
\def\Ib{{\sbm{I}}\XS}

\def\Mb{{\sbm{M}}\XS}  
\def\mb{{\sbm{m}}\XS}

  \def\pb{{\sbm{p}}\XS}
  
  \def\rb{{\sbm{r}}\XS}
  \def\sb{{\sbm{s}}\XS}

  \def\vb{{\sbm{v}}\XS}
  
\def\Xb{{\sbm{X}}\XS}  \def\xb{{\sbm{x}}\XS}
\def\Yb{{\sbm{Y}}\XS}  \def\yb{{\sbm{y}}\XS}
  \def\zb{{\sbm{z}}\XS}

\def\Hb{{\sbm{H}}\XS}  \def\hb{{\sbm{h}}\XS}

 \def\+{^\dagger}

\def\MAP{^{\kern1pt{\rm MAP}\kern-1pt}}

\usepackage{color,soul}

\def\mub         {{\sbmm{\mu}}\XS}

\def\psib        {{\sbmm{\psi}}\XS}        
      
\def\taub        {{\sbmm{\tau}}\XS}

\def\PM{\kern0pt^{\textrm{{\scriptsize PM}}}\kern0pt}
\def\MMAP{\kern1pt^{\textrm{{\tiny MMAP}}}\kern-1pt} 

\def\rem#1{}                    

\captionsetup[subfigure]{subrefformat=simple,labelformat=simple,listofformat=subsimple}

\begin{document}

\maketitle 
\begin{center}
\textbf{Abstract}
\end{center}

\begin{abstract}
M/EEG source localization is an open research issue. To solve it, it is important to have good knowledge of several physical parameters to build a reliable head operator. Amongst them, the value of the conductivity of the human skull has remained controversial. This report introduces a novel hierarchical Bayesian framework to estimate the skull conductivity jointly with the brain activity from the M/EEG measurements to improve the reconstruction quality. A partially collapsed Gibbs sampler is used to draw samples asymptotically distributed according to the associated posterior. The generated samples are then used to estimate the brain activity and the model hyperparameters jointly in a completely unsupervised framework. We use synthetic and real data to illustrate the improvement of the reconstruction. The performance of our method is also compared with two optimization algorithms introduced by Vallagh\'e \textit{et al.} and Gutierrez \textit{et al.} respectively, showing that our method is able to provide results of similar or better quality while remaining applicable in a wider array of situations.
\end{abstract}

\section{Introduction}
M/EEG source localization has been receiving an increasing amount of interest in the signal processing literature in the last decade. One of the most common models used to solve the associated ill-posed problem is the distributed source model that represents the brain activity as a fixed number of dipoles distributed over the brain cortex \cite{grech2008review,hallez2007review}. The relationship between the brain activity and the M/EEG measurements is expressed by the head operator matrix, that is calculated based on the shape and composition of the subject's head. Most of the techniques developed for M/EEG source localization rely on having precise knowledge about the head operator. However there are several factors that introduce uncertainty in its calculation, most noticeably the geometry and conductivity of the different tissues that compose the human head.

\begin{figure}[]
	\centering
	
	\subfloat[][Shell model]{
		\includegraphics{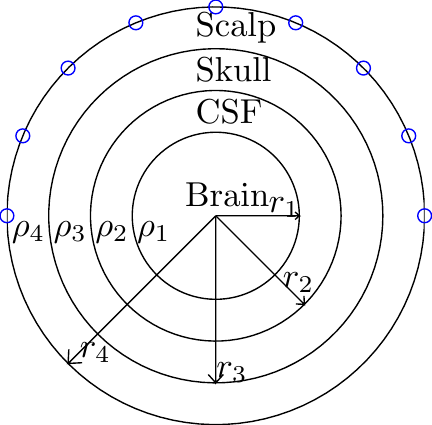}
	}
	\hspace{3em}
	\subfloat[][MRI Scan]{
		\includegraphics[scale=0.35]{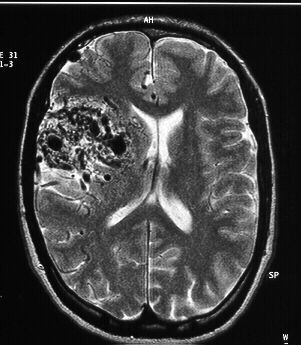}
	}
	\caption{Shell and realistic brain model representations.}
	\label{fig:head_models}	
\end{figure}

Several head models with varying precision and complexity have been used throughout the years, being mainly divided in two categories (1) the shell head models and (2) the realistic head models \cite{hallez2007review} that are represented in Fig. \ref{fig:head_models}. The former model the human head using a fixed number of concentric spheres (typically 3 or 4). Each sphere represents the interface between two different tissues of the human head considered to be uniform with constant conductivities \cite{stok1986inverse}.
In the three-shell head model the skull, cerebrospinal fluid and brain tissues are considered whereas the four-shell model adds an additional outer-most sphere to model the head tissue. In order to calculate the head operator with the shell models it is necessary to set different parameters: (1) the radius of the spheres, (2) the conductivity of each tissue, (3) the amount and locations of the dipoles inside the brain and (4) the amount and locations of the electrodes in the scalp. On the other hand, the realistic head models are typically computed from the MRI of the patient in order to better represent the distribution of different tissues inside the human head. To perform this calculation it is possible to use several methods \cite{hallez2007review} including the boundary element method (BEM) \cite{he1987electric}, the finite element method (FEM) \cite{johnson1995numerical} and the finite difference method (FDM) \cite{marino1993finite}. However, in order to be able to compute a head model operator from the MRI it is still necessary to set several parameters including (1) the conductivity of each tissue, (2) the amount and locations of the dipoles inside the brain and (3) the amount and locations of the electrodes in the scalp.

Since the dipoles are only a discrete approximation of a continuous current distribution, their amount and positions inside the brain are arbitrary \cite{grech2008review}. They are typically distributed uniformly inside the brain and the amount of dipoles is chosen depending on the desired resolution for the activity estimation. In contrast, the other parameters represent physical properties and should be set as close as possible to their real values in order to ensure the quality of the activity estimation.
Considering this, several authors have analyzed the influence of these parameters in the estimation of brain activity. Minor errors in the electrode positions have been shown not to affect significantly the results \cite{wang2001influence} whereas there is a much higher sensibility to variations in the tissue conductivities, making their values critical \cite{acar2013effects,vallaghe2009global,vanrumste2000dipole,genccer2004sensitivity}. The conductivities of the human head tissue, cerebrospinal fluid and brain have well known values that are accepted in the literature \cite{acar2013effects}. However, there has been some controversy regarding the conductivity of the human skull \cite{hallez2007review}. The ratio between scalp and skull conductivities was initially reported to be 80 \cite{geddes1967specific} but since then other authors have published values as low as 15 \cite{hoekema2003measurement}. The value of the human skull conductivity is also known to vary significantly across different subjects \cite{hallez2007review,gonccalves2003vivo}. Because of this, it remains of interest to develop methods that estimate the skull conductivity to improve the quality of the brain activity reconstruction. This can be done using techniques such as impedance tomography (EIT) \cite{gonccalves2003vivo}, using intracranial electrodes \cite{lai2005estimation} or measuring it directly during surgery \cite{hoekema2003measurement}. However it is also possible to estimate it directly from the M/EEG measurements, which is the objective of this paper.

Several methods have been proposed to estimate the conductivity of the skull and the brain activity jointly, albeit requiring very restrictive conditions to yield good results. For instance, having a very good a-priori knowledge about the active dipole positions \cite{gutierrez2004estimating,csengul2012extended}, assuming there is only one dipole active \cite{vallaghe2007vivo} or limiting the estimation of the skull conductivity to a few discrete values \cite{lew2009improved,lew2007low}. In \cite{gutierrez2004estimating} Gutierrez \textit{et al} presented a method capable of estimating the activity amplitude jointly with the skull conductivity as long as the active dipoles are known in advance. The estimation is performed using a concentrated likelihood function that is used to find the maximum likelihood estimator (MLE) of the skull conductivity. Vallagh\'e \textit{et al} \cite{vallaghe2007vivo} introduced an optimization method that estimates the location of one dipole jointly with the conductivities of the different tissues. More precisely, the dipole location and the conductivities were estimated iteratively using the MUSIC algorithm and the Nelder-Mead method.

This paper studies a method able to estimate the skull conductivity jointly with the M/EEG brain activity assuming focal point-like brain activity without requiring any additional prior information. The method is based on a hierarchical Bayesian model introduced in \cite{costa2014l20technicalreport}. This model assigns a multivariate Bernoulli Laplacian prior to the brain activity to approximate an $\ell_{20}$ mixed norm regularization in order to reconstruct point-like brain activity by promoting structured sparsity. In the current work we generalize this model to estimate the skull conductivity jointly with the brain activity in a completely unsupervised framework. The posterior distribution of this generalized model is too complex to derive closed form expressions for the conventional Bayesian estimators of its unknown parameters. Consequently, a Markov chain Monte Carlo method is investigated to generate samples asymptotically distributed according to the posterior distribution of interest. These generated samples are then used to compute Bayesian estimators of the unknown model parameters and hyperparameters.

This paper is organized as follows: The joint EEG deconvolution and operator estimation problem is formulated in Section \ref{sec:problem_statement}. The Bayesian model proposed to solve this problem is introduced in Section \ref{sec:prop_method}. Section \ref{sec:gibbs_sampler} presents the partially collapsed Gibbs sampler that generates samples from the posterior distribution of the proposed model. The method used to model the operator dependency on the skull conductivity is presented in Section \ref{sec:operator_model}. Section \ref{sec:convergence_considerations} studies some interesting moves allowing the sampler convergence to be improved. Section \ref{sec:exp_results} presents experimental results. Section \ref{sec:convergence} investigates how to asses the convergence, and Section \ref{sec:conclusion} finally concludes the paper.
\label{sec:operator_model}

\section{Problem Statement}
\label{sec:problem_statement}
We consider $T$ time samples of $M$ electrodes generated by $N$ dipoles located in the surface of the brain cortex and oriented orthogonally to it. Motivations for this can be found in \cite{grech2008review}. We generalize the model by considering that the operator depends on the skull conductivity
\begin{equation}
\Yb = \Hb (\rho) \Xb + \Eb
\end{equation}
where the dipole amplitudes are represented by $\Xb \in \RR^{N\times T}$, $\Hb (\rho) \in \RR^{M\times N}$ is the lead field operator that varies with the skull conductivity, $\Yb \in \RR^{M\times T}$ contains the electrode measurements and $\Eb \in \RR^{M\times T}$ is the measurement noise.

The problem addressed in this work is the joint estimation of the matrix $\Xb$ and the conductivity $\rho$ from the measurement matrix $\Yb$ using a parametric form for the operator $\Hb$ depending on the conductivity $\rho$.

\subsection{Operator normalization}
\label{subsec:op_norm}
The depth biasing effect is a well known problem when solving the M/EEG source localization by minimizing a certain norm of the solution \cite{grech2008review}. This effect is caused by the fact that each active dipole generates a signal whose amplitude is varying from one dipole to another. It is possible to compensate for this effect by either: (1) adding a weight vector $\vb$ with elements $v_i = |\hb^i|_2$ that penalize the dipoles associated with large M/EEG measurements or (2) normalizing the columns of the operator so that they all have a unitary norm. When using the second option, the EEG source localization problem can be written as

\begin{equation}
\Yb = \bar{\Hb} (\rho) \bar{\Xb} + \Eb
\end{equation}

where the columns of the normalized operator are $\bar{\hb}^i = \frac{\hb^i} {|\hb^i|_2}$ and the rows of the estimated activity are $\bar{\xb}_i = |\hb^i|_2 \xb_i$ being $\mb^i$ the $i$-th column of matrix $\Mb$ and $\mb_i$ its $i$-th row. Since our model has a variable operator, using the first approach would cause the weight vector $\vb$ to depend on the skull conductivity $\rho$ which introduces unnecessary complexities. To avoid these, we use the second compensation model.

\section{Proposed hierarchical Bayesian Model}
\label{sec:prop_method}

\subsection{Likelihood}
We assume an independent white Gaussian noise with a constant variance $\sigma_n^2$, which leads to the following Gaussian likelihood

\begin{equation}
\label{eq:l21_likelihood}
f(\Yb | \bar{\Xb}, \sigma_n^2, \rho) = \prod_{t=1}^T \mathcal{N}\Big(\yb^t \Big| \bar{\Hb} (\rho) \bar{\xb}^t, \sigma_n^2 \II_M\Big)
\end{equation}
where $\II_M$ is the identity matrix of size $M \times M$. Note that if the noise cannot be assumed to be white, it is possible to estimate the noise covariance from the measurements and use it to whiten the data \cite{gramfort2012mixed}.

\subsection{Prior distributions}
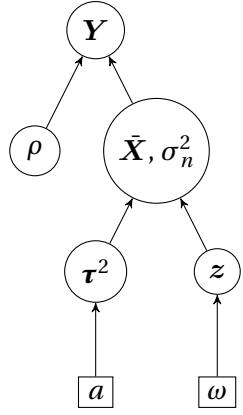
\begin{figure}[H]
\begin{center}
\begin{tikzpicture}[]
   \begin{scope}[
		node distance=1.6cm,on grid,>=stealth',
		hyperpar/.style={rectangle,draw},
		par/.style={circle,draw}]
   \node [par] 	(y)					{$\Yb$};
   \node [par]	(r)[below=of y,xshift=-0.8cm]		{$\rho$}  edge [->] (y);   
   \node [par]	(xs)[below=of y,xshift=0.8cm]		{$\bar{\Xb}$, $\sigma_n^2$}  edge [->] (y);
   \node [par]	(z)	[below=of xs,xshift=0.8cm]	{$\zb$} edge [->] (xs);
   \node [par]	(t) [left=of z]		{$\taub^2$} edge [->] (xs);
   \node [hyperpar]	(a) [below=of t]	{$a$} edge [->] (t);
   \node [hyperpar]	(w) [below=of z]	{$\omega$} edge [->] (z);
   \end{scope}
\end{tikzpicture}
\end{center}   
\caption{Directed acyclic graph for the proposed Bayesian model.}
\label{fig:bayesian_herarchy}
\end{figure}

The dependencies between the different model parameters and hyperparameters after the introduction of the latent variable $\tau^2$ (that simplifies sampling) is shown in Fig. \ref{fig:bayesian_herarchy}. The priors used for each of them (apart from the skull conductivity $\rho$) are the same that were used in the $\ell_{20}$ model of \cite{costa2014l20technicalreport} with $v_i = 1$ (due to the operator normalization described in Section \ref{subsec:op_norm}) and are summarized in Table \ref{tab:table_prior_dist}. The reader is invited to consult \cite{costa2014l20technicalreport} for motivations and more details about our original model. The prior used for the skull conductivity is introduced in Section \ref{subsubsec:skull_cond_prior}.

\begin{table}[H]
	\centering
	\begin{tabular}{|c|l|}
	\hline
	$z_i$ & $
	\mathcal{B}\Big(z_i | \omega\Big)$\\
	\hline
	$\tau_i^2$ & $\mathcal{G}\Big(\tau_i^2 \Big| \frac{T + 1} {2}, \frac{a} {2}\Big)$\\
	\hline
	$\bar{\xb_i}$ & $\begin{array}{ll}
		\delta(\bar{\xb_i}) & \mbox{if } z_i = 0 \\
		\mathcal{N}\Big(0, \sigma_n^2 \tau_i^2 \II_T\Big) & \mbox{if } z_i = 1
	\end{array}$\\
	\hline
	$\sigma_n^2$ & $\frac{1} {\sigma_n^2} 1_{\RR^+} (\sigma_n^2)$\\
	\hline
	$a$ & $\mathcal{G}\Big(a \Big| \alpha, \beta\Big)$\\	
	\hline
	$\omega$ & $\mathcal{U}\Big(\omega \Big| 0, 1\Big)$\\
	\hline
	\end{tabular}
	\caption{Prior distributions $f(z_i | \omega)$, $f(\tau_i^2 | a)$, $f(\bar{\xb_i} | z_i, \tau_i^2, \sigma_n^2)$, $f(\sigma_n^2)$, $f(a)$ and $f(\omega)$.}
	\label{tab:table_prior_dist}
\end{table}
where $1_{\RR^+} (x)$ if $x \in \RR^+$ and 0 otherwise and $\alpha = \beta = 1$.

\subsubsection{Skull conductivity}
\label{subsubsec:skull_cond_prior}
In order to keep the model simple we propose to assign a non-informative uniform prior for the skull conductivity

\begin{equation}
f(\rho) = \mathcal{U}\Big(\rho \Big| \rho_{\min}, \rho_{\max}\Big).
\end{equation}

To select the range of this uniform distribution we consider an interval containing the scalp-to-skull conductivity ratios reported in the literature \cite{geddes1967specific,oostendorp2000conductivity}, i.e, defined by $r_{\min} = 10$ and $r_{\max} = 100$. Considering the scalp conductivity to be $330 \frac{mS} {m}$, this leads to $\rho_{\min} = 3.3 \frac{mS} {m}$ and $\rho_{\max} = 33 \frac{mS} {m}$.

\subsection{Posterior distribution}
The corresponding posterior distribution is defined as follows
\begin{equation}
\label{eq:posterior}
f(\Yb, \sigma_n^2, \bar{\Xb}, \zb, a, \taub^2, \omega, \rho) \propto f(\Yb | \bar{\Xb}, \sigma_n^2, \rho) f(\bar{\Xb} | \taub^2, \zb, \sigma_n^2) f(\zb | \omega) f(\taub^2 | a) f(\sigma_n^2) f(a) f(\omega) f(\rho)
\end{equation}

\section{Partially collapsed Gibbs Sampler}
\label{sec:gibbs_sampler}
Unfortunately, the posterior distribution \eqref{eq:posterior} is intractable and does not allow to express estimators of the different parameters and hyperparameters to be expressed in closed-form. As a consequence, we propose to draw samples from \eqref{eq:posterior} and use them to estimate the brain activity jointly with the model hyperparameters. More precisely, we investigate a partially collapsed Gibbs sampler that samples the variables $z_i$ and $\xb_i$ jointly in order to exploit the strong correlation between these two variables. Denoting as $\Xb_{-i}$ the matrix $\Xb$ whose $i$th row has been replaced by zeros, the proposed sampler samples the different variables according to their conditional distributions as shown in Algorithm \ref{algo:gibbs_sampler}. 

The corresponding conditional distributions are summarized in Table \ref{tab:table_cond_dist} and the next subsection. Their exact derivation can be found in an appendix of \cite{costa2014l20technicalreport}. Note that the only difference between this sampler and the sampler of \cite{costa2014l20technicalreport} is the last line in which the skull conductivity is sampled.

\begin{algorithm}
\caption{Partially Collapsed Gibbs sampler.}
\begin{algorithmic}[]\label{algo:gibbs_sampler}
\STATE  Initialize all the parameters.
\REPEAT
  \FOR {$i=1$ to $N$}
  	\STATE   Sample $\tau_i^2$ from $f(\tau_i^2 | \bar{\xb}_i, \sigma_n^2, a, z_i)$
  	\STATE   Sample $z_i$ from $f(z_i | \Yb, \bar{\Xb}_{-i}, \sigma_n^2, \tau_i^2, \omega, \rho)$
  	\STATE   Sample $\bar{\xb}_i$ from $f(\bar{\xb}_i | z_i, \Yb, \bar{\Xb}_{-i}, \sigma_n^2, \tau_i^2, \rho)$
  \ENDFOR
  \STATE   Sample $a$ from $f(a | \taub^2)$
  \STATE   Sample $\sigma_n^2$ from $f(\sigma_n^2 | \Yb, \bar{\Xb}, \taub^2, \zb, \rho)$
  \STATE   Sample $\omega$ from $f(\omega | \zb)$
  \STATE	   Sample $\rho$ from $f(\rho | \bar{\Xb}, \Yb, \sigma_n^2)$
  \UNTIL {convergence}
\end{algorithmic}
\end{algorithm}

\begin{table}[H]
	\centering
	\begin{tabular}{|c|l|}
	\hline
	$\tau_i^2$ & $
	\begin{array}{ll}
	\mathcal{G}\Big(\frac{T + 1} {2}, \frac{a} {2}\Big) & \mbox{if } z_i = 0 \\
	\mathcal{GIG}\Big(\frac{1} {2}, a, \frac{||\xb_i||^2} {\sigma_n^2}\Big) & \mbox{if } z_i = 1\\
	\end{array}$\\
	\hline
	$z_i$ & $\mathcal{B} \Big(1, \frac{k_1} {k_0 + k_1}\Big)$\\
	\hline
	$\xb_i$ & $\begin{array}{ll}
		\delta(\xb_i) & \mbox{if } z_i = 0 \\
		\mathcal{N}\Big(\mub_i, \sigma_i^2\Big) & \mbox{if } z_i = 1
	\end{array}$\\
	\hline	
	$a$ & $\mathcal{G}\Big(\frac{N(T+1)} {2} + \alpha, \frac{\sum_i{\tau_i^2}} {2} + \beta\Big)$\\
	\hline
	$\sigma_n^2$ & $\mathcal{IG}\Big(\frac{(M+||\zb||_0)T} {2}, \frac{1} {2} \Big[||\bar{\Hb}(\rho) \Xb - \Yb||^2 + \sum_i \frac{||\xb_i||^2} { \tau_i^2}\Big]\Big)$\\
	\hline
	$\omega$ & $\mathcal{B}e\Big(1 + ||\zb||_0, 1 + N - ||\zb||_0\Big)$\\
	\hline
	\end{tabular}
	\caption{Conditional distributions $f(\tau_i^2 | \xb_i, \sigma_n^2, a, z_i)$, $f(z_i | \Yb, \Xb_{-i}, \sigma_n^2, \tau_i^2, \omega, \rho)$, $f(\xb_i | z_i, \Yb, \Xb_{-i}, \sigma_n^2, \tau_i^2, \rho)$, $f(a | \taub^2)$, $f(\sigma_n^2 | \Yb, \Xb, \taub^2, \zb, \rho)$ and $f(\omega | \zb)$.}
	\label{tab:table_cond_dist}
\end{table}
with
\begin{align*}
&\mub_i = \frac{\sigma_i^2 {\bar{\hb}(\rho)^i}^T (\Yb - \bar{\Hb} (\rho) \Xb_{-i})} {\sigma_n^2}, \sigma_i^2 = \frac{\sigma_n^2 \tau_i^2} {1 + \tau_i^2 {\bar{\hb}(\rho)^i}^T \bar{\hb}(\rho)^i}\\
&k_0 = 1 - \omega, 
k_1 = \omega {\left(\frac{\sigma_n^2 \tau_i^2} {\sigma_i^2}\right)}^{-\frac{T} {2}}  \exp{\Big(\frac{||\mub_i||^2} {2 \sigma_i^2}\Big)}.
\end{align*}

\subsection{Conditional distribution of $\rho$}
The following conditional distribution of the skull conductivity can be written

\begin{equation}
f(\rho | \bar{\Xb}, \Yb, \sigma_n^2) \propto \exp\Big(-\frac{||\bar{\Hb} (\rho) \bar{\Xb} - \Yb||^2} {2 \sigma_n^2}\Big) 1_{[\rho_{\min}, \rho_{\max}]} (\rho).
\label{eq:ro_cond_dist}
\end{equation}
For arbitrary functions $\bar{\Hb} (\rho)$ , \eqref{eq:ro_cond_dist} does not belong to a common family of distributions. The following section explains how to efficiently model the operator $\bar{\Hb} (\rho)$ and to sample from \eqref{eq:ro_cond_dist}.

\section{Operator model}
\label{sec:operator_model}
\subsection{Dependency}
\begin{figure}[]
	\includegraphics{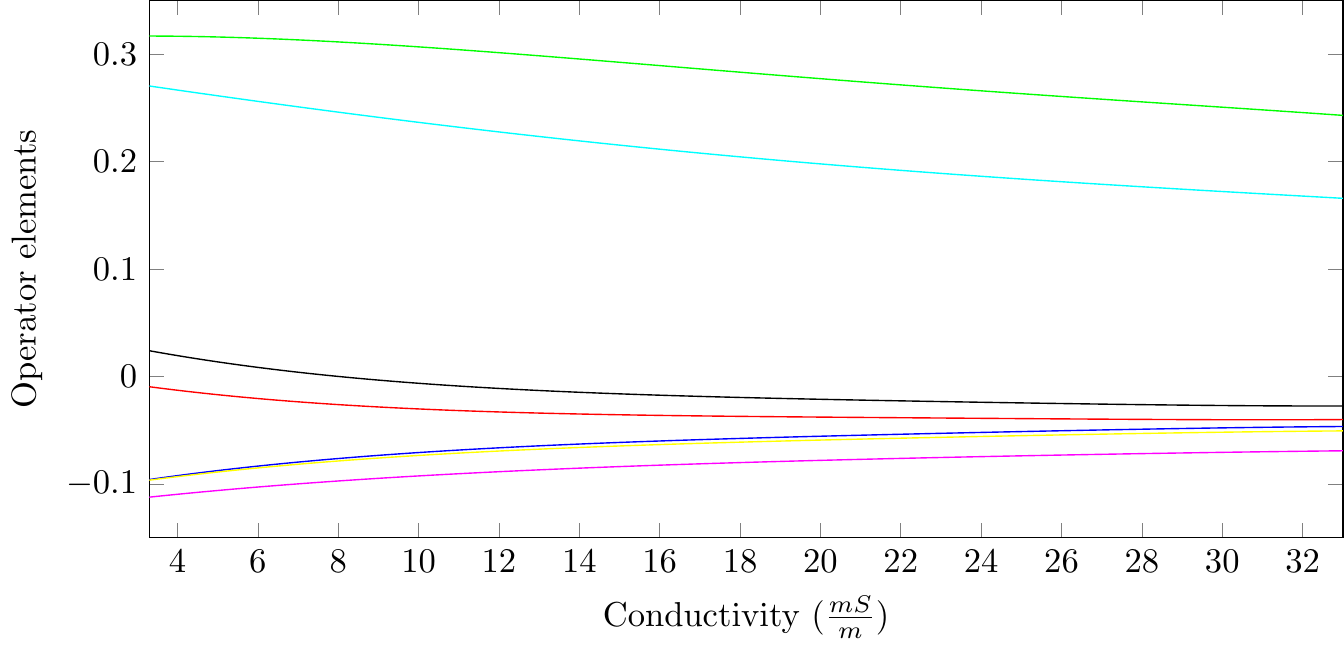}
	\caption{Variations of operator elements with respect to $\rho$.}
	\label{fig:operator_curves}	
\end{figure}

Shell models can be used to derive closed form expressions for $\bar{\Hb} (\rho)$. However these models are quite complex \cite{hallez2007review} and would make the sampling from $f(\rho | \bar{\Xb}, \Yb, \sigma_n^2)$ considerably difficult. In contrast, realistic head models can be calculated numerically for particular values of $\rho$ but do not provide a closed-form expression for $\bar{\Hb} (\rho)$. In order to illustrate how the value of $\rho$ affects the operator, a four-shell 200-dipole head model with 41 electrodes was calculated for different values of $\rho$. Eight elements of the operator $h_{i,j} (\rho)$ (chosen randomly) are displayed in Fig. \ref{fig:operator_curves} as a function of $\rho$. In order to have a simple expression of this dependency, {\c{S}}eng{\"u}l \textit{et al} \cite{csengul2012extended} proposed to replace $\bar{\Hb}(\rho_k)$ by its linearization around the current value $\rho_k$. This method was shown to provide good results but requires evaluating the exact value $\bar{\Hb} (\rho)$ every iteration which slows down the algorithm considerably. Since the variation of the operator elements with respect to $\rho$ is smooth and monotonic, we propose to approximate $\bar{\Hb} (\rho)$ using a simple mathematical expression as in \cite{csengul2012extended}. However, the approximation is computed on the whole range $\rho_{\min} < \rho < \rho_{\max}$ with a polynomial matrix of small degree denoted by $\hat{\Hb} (\rho)$. This allows us to have a simple closed form expression of $\hat{\Hb} (\rho)$ for both shell and realistic head models. In addition, our method only requires calculating the exact $\bar{\Hb}(\rho)$ to construct the polynomial matrix $\hat{\Hb}(\rho)$ offline, after which only the polynomial matrix is evaluated accelerating the iteration speed considerably.

\subsection{Construction of the approximated matrix $\hat{\Hb}(\rho)$ with a polynomial interpolation}
Each element of $\hat{\Hb}(\rho)$ is constructed as a polynomial function of $\rho$

\begin{equation}
\hat{h}_{i,j}(\rho) = \sum_{l=0}^{L} {c_{i,j,l} \rho^l}.
\end{equation}

We propose to calculate the coefficients $c_{i,j,l}$ using polynomial least-squares fitting. This consists in minimizing the following mean square error between $\bar{\Hb} (\rho)$ and $\hat{\Hb} (\rho)$

\begin{equation}
\sum_{k=0}^{K - 1} \Big(\bar{h}_{i,j} (\rho_k) - \sum_{l=0}^L {c_{i,j,l} \rho_k^l}\Big)^2
\label{eq:mse_poly}
\end{equation}

Note that the exact value of $\bar{\Hb} (\rho)$ is calculated for $\rho_k = \rho_{\min} + \frac{k (\rho_{\max} - \rho_{\min})} {K - 1}$ with $0 \leq k \leq {K - 1}$ by either using the shell model expression \cite{stok1986inverse} or by evaluating numerically the head model from the patient's MRI \cite{he1987electric}. After this we calculate the polynomial coefficients $c_{i,j,l}$ that minimize \eqref{eq:mse_poly} that are

\begin{equation}
\cb_{i, j} = \Big[{\sum_{k = 0}^{K - 1}{\psib_k {\psib_k}^T}}\Big]^{-1} \sum_{k = 0}^{K - 1}{\bar{h}_{i,j} (\rho_k) \psib_k}
\label{eq:mse_poly_sol}
\end{equation}

being $\psib_k = [\rho_k^0, ..., \rho_k^L]^T$ and $\cb_{i, j} = [c_{i, j, 0}, ..., c_{i, j, L}]^T$.

The next section will discuss how to sample the skull conductivity after the coeficients $c_{i, j, l}$ have been calculated.
	
\subsection{Sampling the skull conductivity}
Approximating the relationship between the operator and the skull conductivity with a polynomial matrix allows us to have a simple closed form expression for $f(\rho | \bar{\Xb}, \Yb, \sigma_n^2)$

\begin{equation}
f(\rho | \bar{\Xb}, \Yb, \sigma_n^2) \propto \exp\Big(-g(\rho)\Big) 1_{[\rho_{\min}, \rho_{\max}]} (\rho)
\label{eq:cond_ro_approx}
\end{equation}

where $g(\rho) = \frac{||\hat{\Hb}_{L,K}(\rho) \bar{\Xb} - \Yb||^2} {2 \sigma_n^2}$ is a polynomial of order $2 L$.

Since it is not easy to sample from \eqref{eq:cond_ro_approx} directly, we propose to adopt a random-walk Metropolis-Hastings (MH) move. More precisely, this move consists in proposing a new sample $\rho_{\textrm{prop}} = \rho_{\textrm{old}} + \epsilon$ and accepting it with probability

\begin{equation}
P_{acc} = 
\left\{
	\begin{array}{ll}
		\min\Big(\frac{\exp(-g(\rho_{prop}))} {\exp(-g(\rho_{old}))}, 1\Big)  & \mbox{if } \rho_{\min} < \rho_{\textrm{prop}} < \rho_{\max}\\
		0 & \mbox{otherwise}.
	\end{array}
\right.
\end{equation}

We propose to use a zero-mean Gaussian distribution for the random walk noise $\epsilon$, i.e., $f(\epsilon) = \mathcal{N}(0, \sigma_{\epsilon}^2)$. The variance of the random walk $\sigma_{\epsilon}^2$ was adjusted empirically in order to obtain an appropriate acceptance rate, as recommended in \cite{casella1999monte}. Based on our experiments performed in Section \ref{sec:exp_results}, we set $\sigma_{\epsilon} = \frac{\rho_{\max} - \rho_{\min}} {100}$.

\section{Convergence diagnosis}
\label{sec:convergence_considerations}
\subsection{Multiple dipole shift proposals}

As noticed in our previous work \cite{costa2014l20technicalreport}, the Gibbs sampler presented above tends to gets stuck around local maxima of the posterior distribution. In particular, the variable $\zb$ does not usually converge to indicate the correct positions of the active dipoles in a reasonable amount of time.
To solve this problem we introduced in \cite{costa2014l20technicalreport} multiple dipole shift proposals allowing $D$ random non-zeros to be moved to neighboring positions. These moves are accepted or rejected using the classical MH acceptance rate.

\begin{algorithm}
\caption{Multiple dipole shift proposal.}
\begin{algorithmic}[!H]\label{algo:proposal}
\STATE   $\hat{\zb} = \zb$
  \STATE \textbf{repeat} D times
  	\STATE\hspace{\algorithmicindent}	Set $\textrm{ind}_{\textrm{old}}$ to be the index of a random non-zero of $\zb$
  	\STATE\hspace{\algorithmicindent}	Set $\pb = [\textrm{ind}_{\textrm{old}}, \textrm{neigh}_{\gamma,\rho}(\textrm{ind}_{\textrm{old}})]$
  	\STATE\hspace{\algorithmicindent} 	Set $\textrm{ind}_{\textrm{new}}$ to be a random element of $\pb$ 
  	\STATE\hspace{\algorithmicindent}	Set $\hat{z}_{\textrm{ind}_{\textrm{old}}} = 0$ and $\hat{z}_{\textrm{ind}_{\textrm{new}}} = 1$
  \STATE \textbf{end}
  	\STATE	Sample $\hat{\rho}$ from $f(\hat{\rho} | \Yb, \sigma_n^2, \taub^2, \hat{\zb})$.  
  	\STATE	Sample $\hat{\Xb}$ from $f(\hat{\Xb} | \hat{\zb}, \Yb, \sigma_n^2, \taub^2, \hat{\rho})$.    
  	\STATE	Sample $\hat{\taub}^2$ from $f(\hat{\taub}^2 | \hat{\Xb}, \sigma_n^2, a, \hat{\zb})$.
  \STATE		Set $\{\zb,\taub^2, \rho\} = \{\hat{\zb}, \hat{\taub}^2, \hat{\rho}\}$ with probability $\min\Big(\frac {f(\hat{\zb}, \hat{\taub}^2, \hat{\rho} | \Yb, a, \sigma_n^2, \omega)} {f(\zb, \taub^2, \rho | \Yb, a, \sigma_n^2, \omega)}, 1\Big)$
  \STATE    Else, do not change the values of $\{\zb, \tau^2, \rho\}$
  \STATE		Resample $\Xb$ if the proposed move has been accepted
\end{algorithmic}
\end{algorithm}

In order to build more efficient proposals, we expand the multiple dipole shift moves to include the skull conductivity $\rho$. We do this by adding a step to sample the skull conductivity $\rho$ marginally to $\Xb$. The final algorithm is illustrated in Algorithm \ref{algo:proposal} where we define the neighborhood of each dipole as
\begin{equation}
\label{eq:neigh_def}
\textrm{neigh}_{\gamma,\rho}(i) \triangleq \{j \neq i \enspace|\enspace |\textrm{corr}({\hat{\hb}(\rho)}^i, {\hat{\hb}(\rho)}^j)| \geq \gamma\}
\end{equation}
where $\hat{h}(\rho)^i$ is the $i$th column of $\hat{\Hb}(\rho)$, $\gamma \in [0,1]$ tunes the neighborhood size ($\gamma = 0$ corresponds to a neighborhood containing all the dipoles and $\gamma = 1$ corresponds to an empty neighborhood). The values of $D$ and $\gamma$ were adjusted by cross validation to $D = 2$ and $\gamma = 0.8$.


To accept or reject the moves introduced before, we need to evaluate the distribution $f(\zb, \taub^2, \rho | \Yb, a, \sigma_n^2, \omega)$ up to a multiplicative constant. Considering only the values of the subset of $\zb$ that change between the current value and the proposal (denoted by $\rb$) and applying the arithmetic manipulations detailed in \cite{costa2014l20technicalreport}, the following result can be obtained

\begin{equation}
f(\zb_r, \taub_r^2, \rho | \Yb, a, \sigma_n^2, \omega) \propto (1-\omega)^{C_0} \omega^{C_1} (\sigma_n^2)^{-\frac{T C_1} {2}} |\Sigma|^{\frac{T} {2}} \prod_{i \in \Ib_1} {(\tau_i^2)^{-\frac{T} {2}}} \exp\Big({-\frac{\sum_{t=1}^T Q^t} {2}}\Big) \prod_{i = 1}^{N}{\mathcal{G}\Big(\tau_i^2 \Big| \frac{T + 1} {2}, \frac{a} {2}\Big)}
\label{eq:acc_prob}
\end{equation}

where $\Ib_j = \{i | z_{r_i} = j\}$, $C_j$ is the cardinal of the set $\Ib_j$ (for $j = \{0, 1\}$) and

\begin{align*}
\Sigma^{-1} &= {\frac{1} {\sigma_n^2}\left[{\Hb^{\Ib_1}(\rho)^T \Hb^{\Ib_1}(\rho) + \diag\Big(\frac{1} {\tau^2_{\rb}}\Big)}\right]}\\
\mu^t &= -\frac{\Sigma \Hb^{\Ib_1}(\rho)^T [\Hb^{-\rb}(\rho) \xb_{-\rb}^t - \yb^t]} {\sigma_n^2}\\
Q^t &= \frac{[\Hb^{-\rb}(\rho) \xb_{-\rb}^t - \yb^t]^T [\Hb^{-\rb}(\rho) \xb_{-\rb}^t - \yb^t]} {\sigma_n^2} - {\mu^t}^T  \Sigma^{-1} \mu^t
\end{align*}

with $\diag(\sb)$ the diagonal square matrix formed by the elements of the vector $\sb$.

\subsection{Inter-chain proposals}
As explained in our previous work \cite{costa2014l20technicalreport}, when running  multiple MCMC chains in parallel it is possible for the different chains to get stuck around different values of $\zb$. As a consequence, inter-chain proposals can be considered to improve the sampler convergence. These proposals were exchanging active dipoles of different chains, with a probability defined according to the MH acceptance rate.

\begin{algorithm}
\caption{Inter-chain proposals.}
\begin{algorithmic}[!H]\label{algo:inter_proposal}
\STATE   Define a vector $\cb = \{1, ..., L\}$ where L is the number of chains
  \STATE \textbf{for} $i = \{1, 2, ..., L\}$
  	\STATE\hspace{\algorithmicindent}	Choose (and remove) a random element from $\cb$ and denote it by $k$
	\STATE\hspace{\algorithmicindent} Denote as $\{\hat{\zb}_k, \hat{\taub}_k^2, \hat{\rho}_k\}$ the sampled values of $\{\zb, \taub^2, \rho\}$ for the MCMC chain $\#k$
	  \STATE	\hspace{\algorithmicindent} 	For the chain $\#i$ set $\{\zb_i,\taub_i^2,\rho_i\} = \{\hat{\zb}_k,\hat{\taub}_k^2,\hat{\rho}_k\}$ with probability $\frac {f({\hat{\zb}_k}, \hat{\taub}_k^2, \hat{\rho}_k | .)} {f(\zb, \taub^2, \rho | .)}$
  \STATE\hspace{\algorithmicindent}		Resample $\Xb$ if the proposal has been accepted
  \STATE \textbf{end}
\end{algorithmic}
\end{algorithm}

We now expand the inter-chain proposals to include the conductivity $\rho$ to improve the sampling efficiency. More precisely, at each iteration, an exchange between the active dipoles of random pairs of chains is proposed with probability $\pi$ (the value of $\pi$ was fixed to $10^{-3}$ by crossvalidation) and this exchange is accepted with the MH probability as shown in Algorithm \ref{algo:inter_proposal}.

\section{Experimental Results}
\label{sec:exp_results}

\subsection{Synthetic data}
Synthetic data is first considered using a 200-dipole Stok four-sphere head model \cite{stok1986inverse} with 41 electrodes. The dipoles were uniformly distributed in the brain cortex oriented orthogonally to the brain surface. Two different kinds of activations are investigated: (1) single dipole activations with low SNR and (2) multiple dipole activations with high SNR. Before applying our method to the data we will evaluate the sensitivity of the method to the values of $K$ and $L$.

\subsubsection{Selection of K and L}
\label{subsec:set_p_and_p}

In order to construct $\hat{\Hb}(\rho)$ we have to select appropriate values for $K$ and $L$. Increasing the values of $K$ and $L$ clearly improves the quality of the approximation of $\hat{\Hb}$ at the price of higher computational complexity. To choose appropriate values for $K$ and $L$, it is interesting to analyze their effect on the estimation of $\rho$ when all the other parameters ($\Xb$ and $\sigma_n^2$) are known. To do this we will use several ground truth values of $\Xb$ and $\sigma_n^2$ and consider that the algorithm would ideally converge to the value $\hat{\rho}$ that minimizes the polynomial of order $2 L$ defined as $g(\rho) = \frac{||\hat{\Hb}_{L,K}(\rho) \bar{\Xb} - \Yb||^2} {2 \sigma_n^2}$.

Using the head model described in the previous subsection, we used $K = 100$ and generated values of $L$ in the range $2 \leq L \leq 7$. We also considered ten different ground truth values of $\rho_{gt}$, 200 different values of $\bar{\Xb} \in \RR^{200 \times 100}$ (each of them having one of the $200$ dipoles with a constant activity during the $100$ time samples) and 9 values of SNR (from $0$dB to $40$dB in steps of $5$dB) resulting in a total of 108.000 experiments. Since the ground truth $\bar{\Xb}$ is used for the estimation of $\hat{\rho}$, the only factors that explain the difference between $\hat{\rho}$ and $\rho_{gt}$ are the presence of noise and the approximation of $\bar{\Hb}(\rho)$ by $\hat{\Hb}_L(\rho)$, which allows us to illustrate the effect of $L$ for different values of SNR. We define the root mean square error of the estimation of $\rho$ for a particular value of $L$ and $SNR$ as

\begin{equation}
RMSE(L, SNR) = \sqrt{\frac{1} {200 \times 10} \sum_{i=1}^{200 \times 10} (\widehat{\rho}_i - \rho_i)^2}
\end{equation}
$\rho_i$ being the ground truth of $\rho$ for the $i$th realization obtained for the specified values of $L$ and $SNR$ and $\hat{\rho}_i$ the corresponding estimated value of $\rho$.

\begin{figure}[]
	\centering
	\subfloat[][RMSE of $\rho$ VS L (K = 100). Noiseless (left) and noisy (right) measurements.]{
		\label{fig:cond_error_pol_order}	\includegraphics{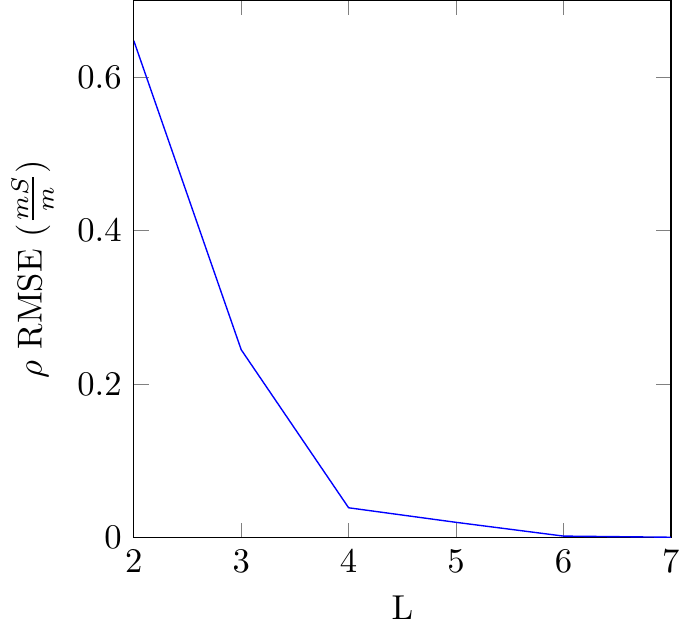}	\includegraphics{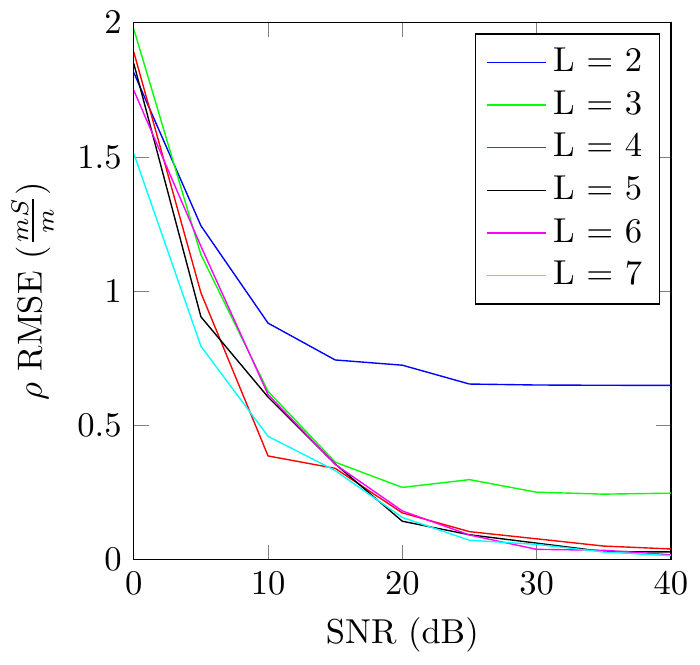}
	}
	
	\subfloat[][RMSE of $\rho$ VS K (L = 7). Noiseless (left) and noisy (right) measurements.]{	
		\label{fig:cond_error_int_points}	\includegraphics{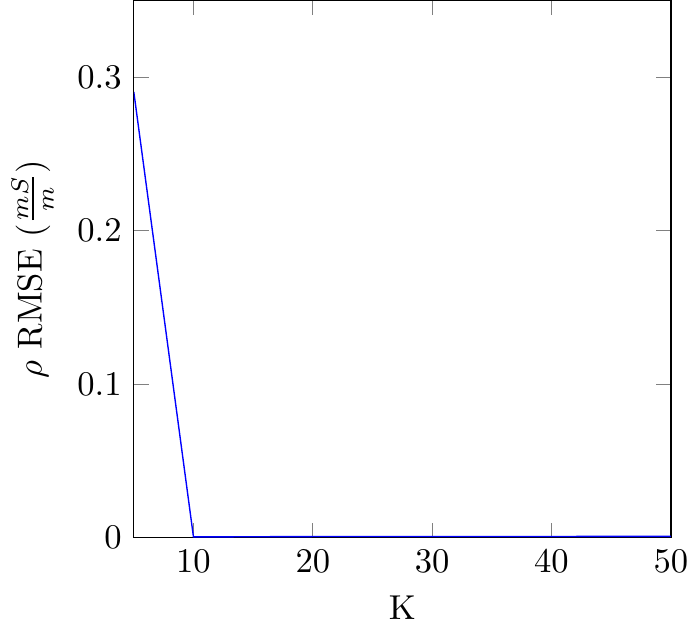}
	\includegraphics{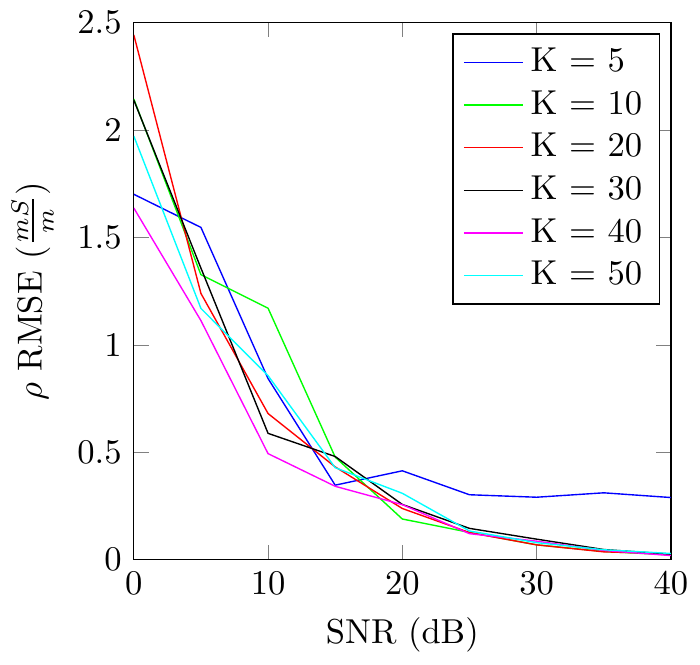}
	
			}
	\caption{Root mean square error of the skull conductivity estimation}
\end{figure}

Fig. \subref*{fig:cond_error_pol_order} illustrates the RMSE of the estimation of $\rho$ as a function of SNR for each value of $L$ and for noiseless simulations. We can see that in noiseless situations, the error seems to tend asymptotically to 0 as the value of $L$ increases, as expected. However, when the measurements are noisy the minimum value of RMSE is limited by the amount of noise. For instance, if the measurements have an SNR of $10$dB the estimation error only decreases until $L = 3$. This shows that for high SNR it makes sense to choose a high value of $L$ but for common values of SNR (lower than $20$dB) choosing high values of $L$ does not improve the quality of the estimation of $\rho$. Because of this, for a given SNR, we choose the smallest value of $L$ such that $|RMSE(SNR, L + 1) - RMSE(SNR, L)| < \epsilon$.

To analyze the effect of $K$, the same experiment was performed by varying $K \in \{5, 10, 20, 30, 40\}$ and fixing $L = 5$ for $K = 5$ and $L = 7$ for the higher values of $K$. Fig. \subref*{fig:cond_error_int_points} shows the RMSE of $\rho$ for noisy and noiseless simulations. Both figures show that there is a significant improved approximation when $K$ is increased from $5$ to $10$ if the SNR is high enough. However, choosing values of $K$ higher than $10$ does not improve the estimation of the skull conductivity $\rho$. Since we will be working with values of SNR between $10$ and $30$dB we have decided to use $L = 4$ and $K = 10$ in the rest of the paper.

\begin{figure}[]
	\centering
	\subfloat[][Histograms of $a$, $\omega$, $\sigma_n^2$ and $\rho$ for the proposed model ($\rho$ estimated).]{
		\includegraphics{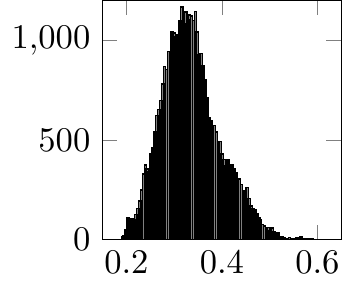}	\includegraphics{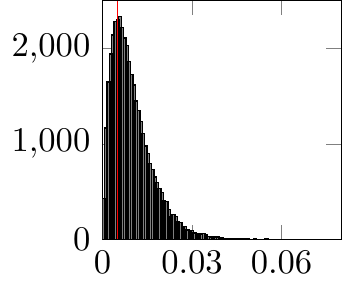}	\includegraphics{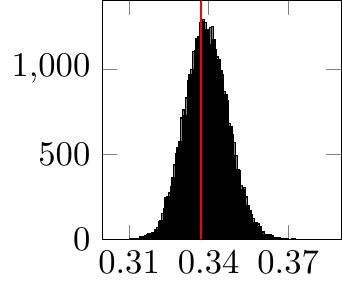}
\includegraphics{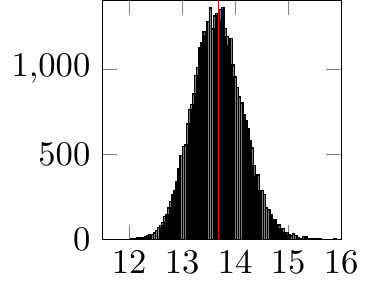}
	}

	\subfloat[][Histograms of $a$, $\omega$ and $\sigma_n^2$ for the default-$\rho$ model ($\rho = \rho_{fix}$ is not estimated)]{	\includegraphics{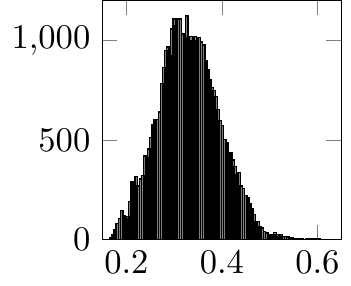}	\includegraphics{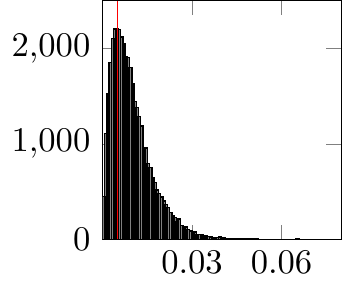}	\includegraphics{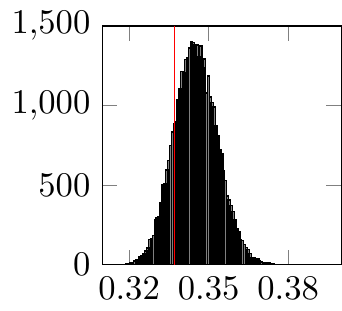}
	}
	\caption{Estimated posterior distribution of the model parameters for simulation $\#1$ (single dipole)}
	\label{fig:single_dipole_2_histograms}
\end{figure}

\subsubsection{Single dipole activations}
\label{subsec:single_dipole}
In the first kind of simulations, a random single dipole was assigned a damped sine activation of $5$Hz. The activation was sampled at $200$Hz and multiplied with the head operator with a chosen ground truth value of $\rho$ denoted as $\rho_{gt}$. Gaussian white noise was added to the measurements to have a signal to noise ratio $SNR = 10$dB. The proposed estimation method is compared with two other methods: (1) a variation of the proposed method that uses a fixed value of the conductivity, more precisely $\rho_{fix} = \frac{\rho_{\max} + \rho_{\min}} {2} = 18.15 \frac{mS} {m}$ in order to illustrate the advantages of estimating $\rho$ (called default-$\rho$ model) and (2) the optimization method studied by Vallagh\'e \textit{et al} \cite{vallaghe2007vivo} that is able to estimate $\rho$ and the brain activity jointly if there is only one active dipole. Both the proposed model and the default-$\rho$ model use 8 parallel MCMC chains. We illustrate two different cases. The first case corresponds to a conductivity $\rho_{gt}$ close to $\rho_{fix}$ (simulation $\#1$) and the second case is characterized by a higher difference between $\rho_{gt}$ and $\rho_{fix}$ (simulation $\#2$). More precisely, $\rho_{gt} = 13.68 \frac{mS} {m}$ for simulation $\#1$ and $\rho_{gt} = 3.59 \frac{mS} {m}$ for simulation $\#2$.

\begin{figure}[]
	\centering	\includegraphics{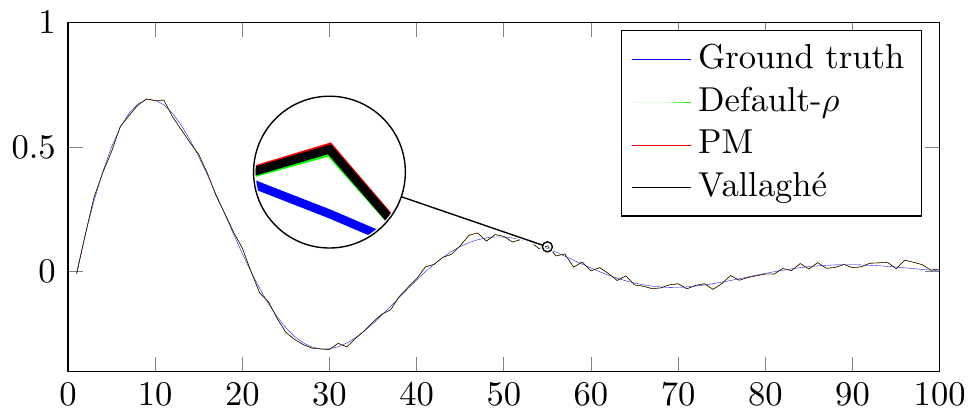}

	\caption{Recovered waveforms in  the single dipole simulation $\#$1.}
	\label{fig:waveforms_single_dipole_2}	
\end{figure}

\begin{figure}[]
	\centering
	\subfloat[][Ground truth - Axial, coronal and sagittal views respectively]{
		\includegraphics{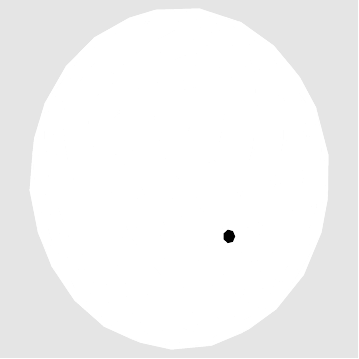}
		\includegraphics{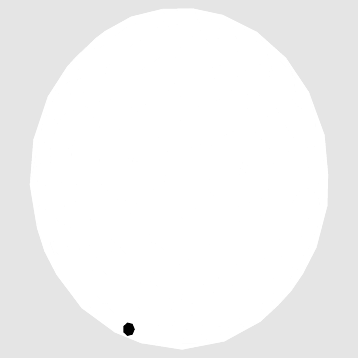}
 		\includegraphics{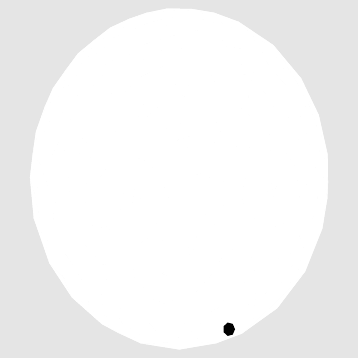}
	}

	\subfloat[][Default-$\rho$ model - Axial, coronal and sagittal views respectively]{
		\includegraphics{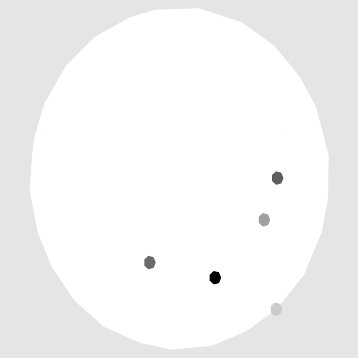}
		\includegraphics{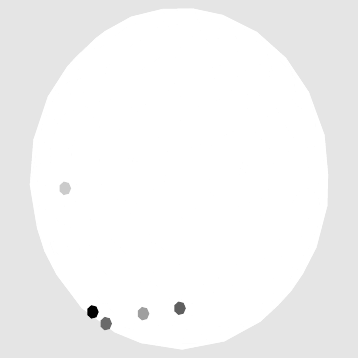}
		\includegraphics{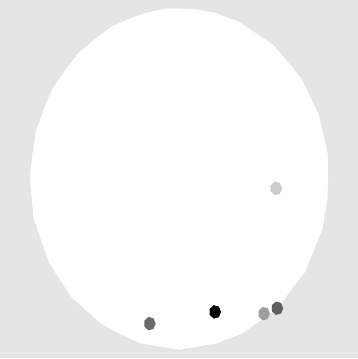}
	}
	
	\subfloat[][Proposed method - Axial, coronal and sagittal views respectively]{
		\includegraphics{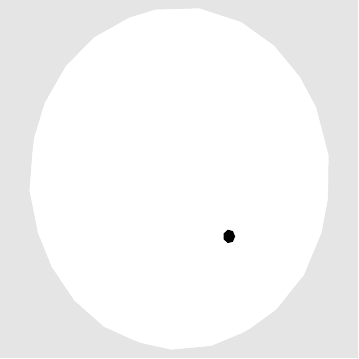}
		\includegraphics{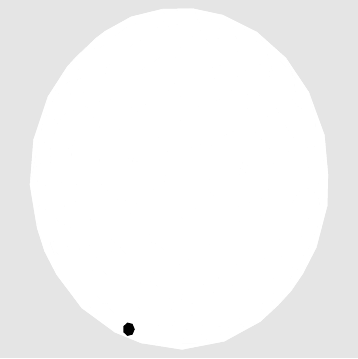}
 		\includegraphics{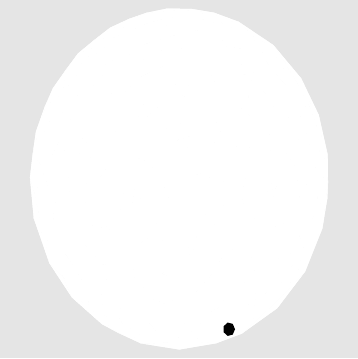}
	}
	
	\subfloat[][Vallagh\'e's method - Axial, coronal and sagittal views respectively]{
		\includegraphics{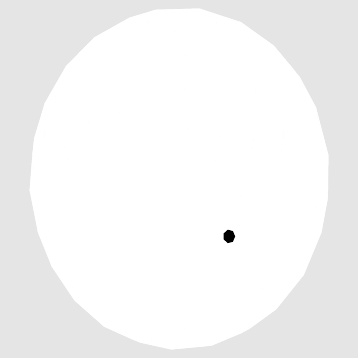}
		\includegraphics{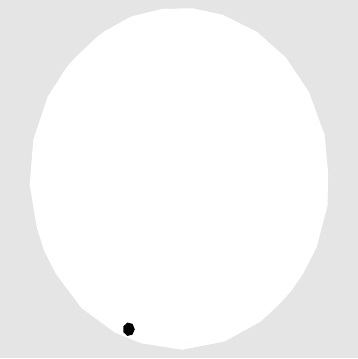}
 		\includegraphics{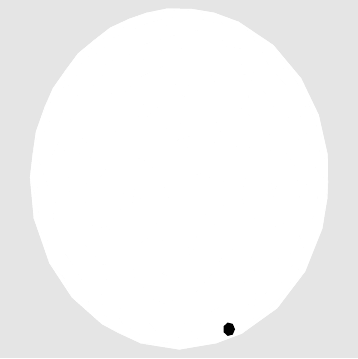}
	}	
	\caption{Estimated activity for single dipole simulation $\#2$.}
	\label{fig:single_dipole_location_1}
\end{figure}

\begin{figure}[!]
	\centering
	\includegraphics{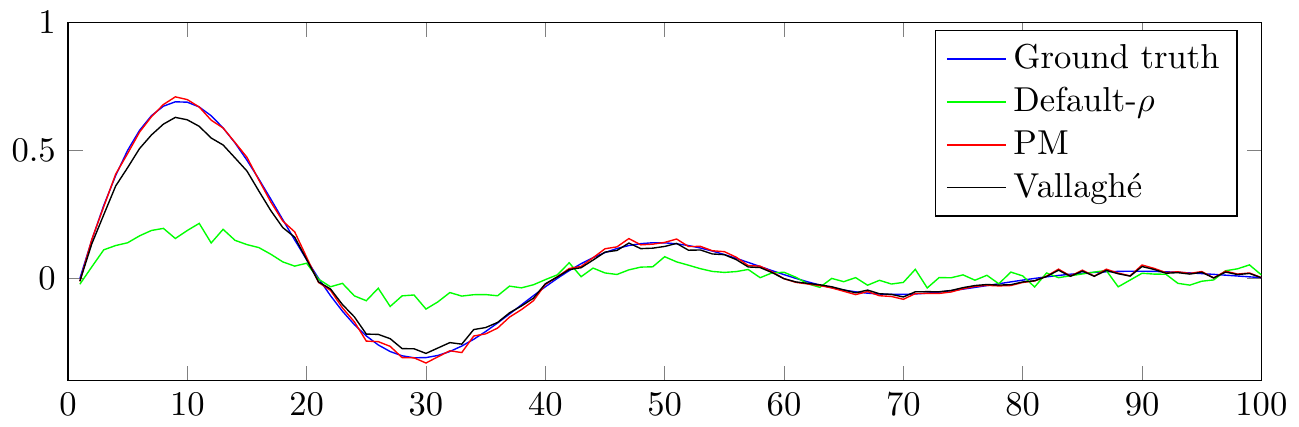}
	\caption{Estimated waveforms for simulation $\#$2 (single dipole)}
	\label{fig:waveforms_single_dipole_1}	
\end{figure}	

For simulation \#1, all methods manage to find the active dipole in its exact location. The histograms of the simulated parameters for the two Bayesian models (corresponding to a fixed value of $\rho$ and to an unknown value of $\rho$ respectively) are shown in Fig. \ref{fig:single_dipole_2_histograms}. First we comment the results obtained with the proposed model and its default-$\rho$ variation. In order to understand why the model with a fixed value of $\rho$ is able to correctly recover the activation despite the use of a wrong operator, it is important to note on the different histograms that the proposed model is able to estimate all variables correctly. Note also that the default-$\rho$ model seems to overestimate the noise variance $\sigma_n^2$. Thus, when the difference between the operators is small, the default-$\rho$ model is able to mitigate the effect of a wrong operator by considering that there is additional noise in the system.

Vallagh\'e's optimization method also estimates the dipole location correctly for simulation $\#1$. After running simulation $\#1$ for 20 Monte Carlo runs with different values of noise, the mean estimate value of the skull conductivity was $\hat{\rho} = 13.63 \frac{mS} {m}$ (which is very close to the actual ground truth value $13.69 \frac{mS} {m}$). The proposed method is able to estimate the posterior distribution of the skull conductivity (as shown in Fig. \ref{fig:single_dipole_2_histograms}). This distribution can be used to determine the MMSE estimator of $\rho$ (the mean of the posterior distribution, that is $13.64 \frac{mS} {m}$) as well as uncertainties regarding this estimator. For instance the standard deviation of the MMSE estimator is $0.46 \frac{mS} {m}$. Since the proposed method is not restricted to a point-estimate as Vallagh\'e's optimization method, it is not necessary to run different Monte Carlo runs for different values of noise.

Finally, Fig. \ref{fig:waveforms_single_dipole_2} compares the estimated waveforms obtained with the proposed method (PM) and Vallagh\'es one, where we can observe that both estimated waveforms are close to the ground truth. Quantitative results associated with simulation $\#$1 are summarized in Table \ref{fig:single_dipole_1_results}.

The estimated locations for simulation $\#2$ are shown in Fig. \ref{fig:single_dipole_location_1}. In this case the default-$\rho$ model fails to recover the correct dipole location and spreads the activity over a significant area of the brain, due to the fact that the difference between the operators is significantly higher. The optimization method is still able to recover the dipole position correctly and gives an average estimate of the skull conductivity $\hat{\rho} = 3.85 \frac{mS} {m}$ over 20 Monte Carlo runs while the proposed method estimates a mean value of $3.49 \frac{mS} {m}$ (closer to the ground truth value of $3.59 \frac{mS} {m}$) with a standard deviation of $0.12 \frac{mS} {m}$. Fig. \ref{fig:waveforms_single_dipole_1} shows that in this case the proposed method estimates a waveform that is considerably closer to the ground truth. Table \ref{fig:single_dipole_2_results} summarizes quantitative results of simulation $\#$2.

The price to pay with the good performance of the proposed method is its computational complexity. Using Matlab implementations in a modern Xeon CPU E3-1240 processor, each simulation was processed on average in $96.1$ seconds by the proposed model, $51.29$ seconds by the default-$\rho$ model and $23.8$ seconds by Vallagh\'e's method. This is a common disadvantage of Bayesian methods when compared with optimization techniques.

In summary, both the proposed and Vallagh\'e's method are able to find the correct dipole position for both simulations but the proposed method is able to better estimate the skull conductivity and the activation waveforms for simulation \#2 while not being restricted to single dipole activations as Vallagh\'e's method.

\begin{figure}[t]
	\centering
	\subfloat[][Recovery rate versus C.]{
		\label{fig:rec_rate}
		\includegraphics{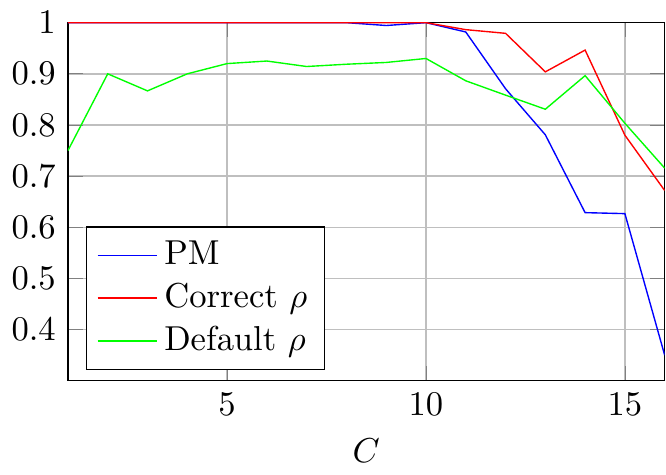}
	}
	\subfloat[][Proportion of residual energy versus C.]{
		\label{fig:res_energy}
		\includegraphics{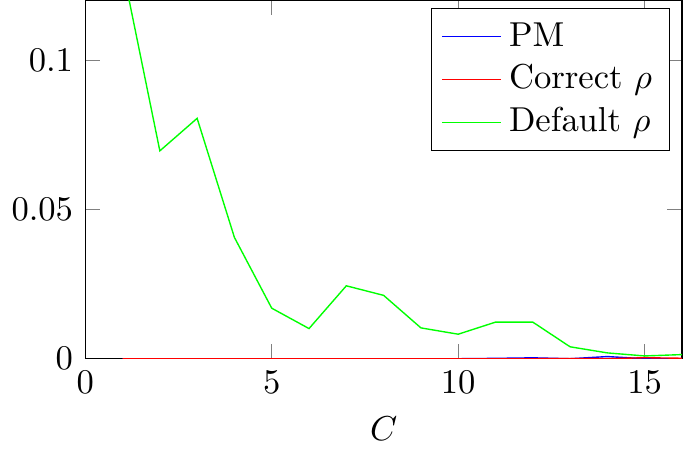}
	}
	
	\subfloat[][Root mean square error of $\hat{\rho}$ estimation]{
		\label{fig:pcee}
		\includegraphics{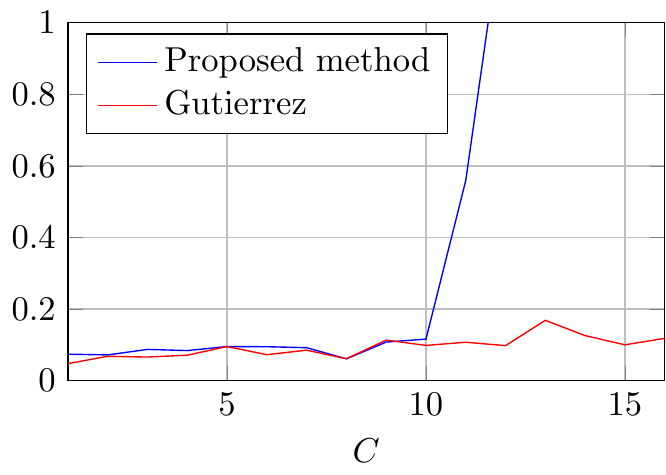}
	}
	\caption{Performance measures for the estimation of multiple dipoles as a function of $C$.}
	\label{fig:multiple_dipole_recovery_rate}
\end{figure}

\begin{table}[]
	\centering
	\begin{tabular}{|c|c|c|c|c|c|}
	\hline 
	Method & Position error & $\frac{|\hat{X} - X_{gt}|_F^2} {|X_{gt}|_F^2}$ & $\frac{\hat{\sigma_n}^2 - \sigma_n^2} {\sigma_n^2}$ & $\hat{\rho} - \rho_{gt}$ & $\frac{|\Hb(\rho_{gt}) - \Hb(\hat{\rho})|_F^2} {|\Hb(\rho_{gt})|_F^2}$\\
	\hline
	Default-$\rho$ & $0$ & $2.53 \times 10^{-3}$ & $26.99 \times 10^{-3}$ & $4.46 \times 10^{-3}$ & $2.51 \times 10^{-3}$\\	
	PM & $0$ & $2.47 \times 10^{-3}$ & $6.21 \times 10^{-3}$ & $-43.89 \times 10^{-6}$ & $1.21 \times 10^{-6}$\\	
	Vallagh\'e & $0$ & $128 \times 10^{-6}$ & N/A & $-52.83 \times 10^{-6}$ & $1.15 \times 10^{-6}$\\
	\hline
	\end{tabular}
	\caption{Estimation errors for the different parameters (Simulation \#1)}
	\label{fig:single_dipole_1_results}
\end{table}

\begin{table}[]
	\centering
	\begin{tabular}{|c|c|c|c|c|c|}
	\hline 
	Method & Position error & $\frac{|\hat{X} - X_{gt}|_F^2} {|X_{gt}|_F^2}$ & $\frac{\hat{\sigma_n}^2 - \sigma_n^2} {\sigma_n^2}$ & $\hat{\rho} - \rho_{gt}$ & $\frac{|\Hb(\rho_{gt}) - \Hb(\hat{\rho})|_F^2} {|\Hb(\rho_{gt})|_F^2}$\\
	\hline
	Default-$\rho$ & $224 \times 10^{-3}$ & $1.36$ & $108.4 \times 10^{-3}$ & $14.56 \times 10^{-3}$ & $75.85 \times 10^{-3}$\\	
	PM & $0$ & $2.47 \times 10^{-3}$ & $5.89 \times 10^{-3}$ & $10.33 \times 10^{-6}$ & $77.14 \times 10^{-9}$\\	
	Vallagh\'e & $0$ & $6.17 \times 10^{-3}$ & N/A & $265 \times 10^{-6}$ & $92.3 \times 10^{-6}$\\
	\hline
	\end{tabular}
	\caption{Estimation errors for the different parameters (Simulation \#2)}
	\label{fig:single_dipole_2_results}
\end{table}

\subsubsection{Deep dipole activations}
In order to investigate how the algorithm performs for active dipoles at different depths, the following set of experiments was performed:

The original 200-dipole brain model was expanded to have 600 dipoles. The first 200 were localized in the same positions as in the original experiment (henceforth called ``superficial dipoles"), the second set of 200 dipoles were localized 10\% closer to the center of the sphere model compared to the first 200 dipoles (henceforth called ``medium dipoles"). The final 200 dipoles were located 20\% closer to the center of the sphere model than the superficial dipoles, and will be referred to as ``deep dipoles".

With this new brain model, a random superficial dipole was chosen. This superficial dipole was assigned a damped sinusoidal brain activity wave. In two separate scenarios, this activation wave was moved to the closest medium dipole and the closest deep dipole respectively, resulting in three values of $\bold{X}$. For each of these three activations 10 linearly spaced values of  ground truth $\rho$ were chosen to calculate the leadfield matrix $\bold{H}$, resulting in 30 noiseless measurements $\bold{H X}$. Finally, noise was added to each of these noiseless measurements in order to obtain different values of the SNR, i.e., SNR = 10, 15, 20 and 30dB, resulting in a total of 120 noisy measurements $\bold{Y}$. The proposed method was applied to each of the values of $\bold{Y}$ to recover the original value of $\bold{X}$ jointly with $\rho$.

\begin{figure}[]
	\centering
	\subfloat[][Superficial dipole]{\label{fig:sk_cond_est1}
		\includegraphics[width=150pt, height=150pt]{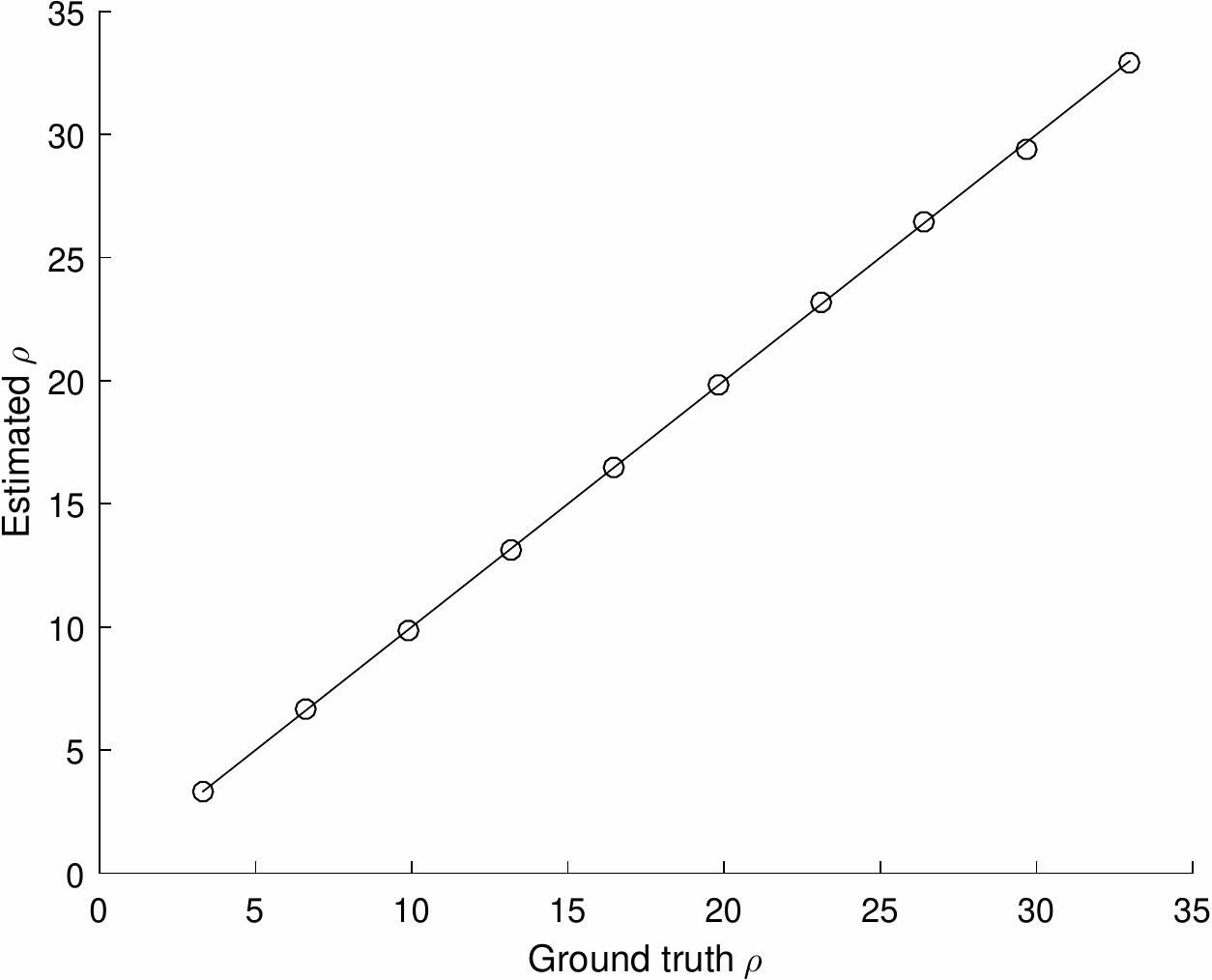}
	}
	\subfloat[][Medium dipole]{\label{fig:sk_cond_est2}	
		\includegraphics[width=150pt, height=150pt]{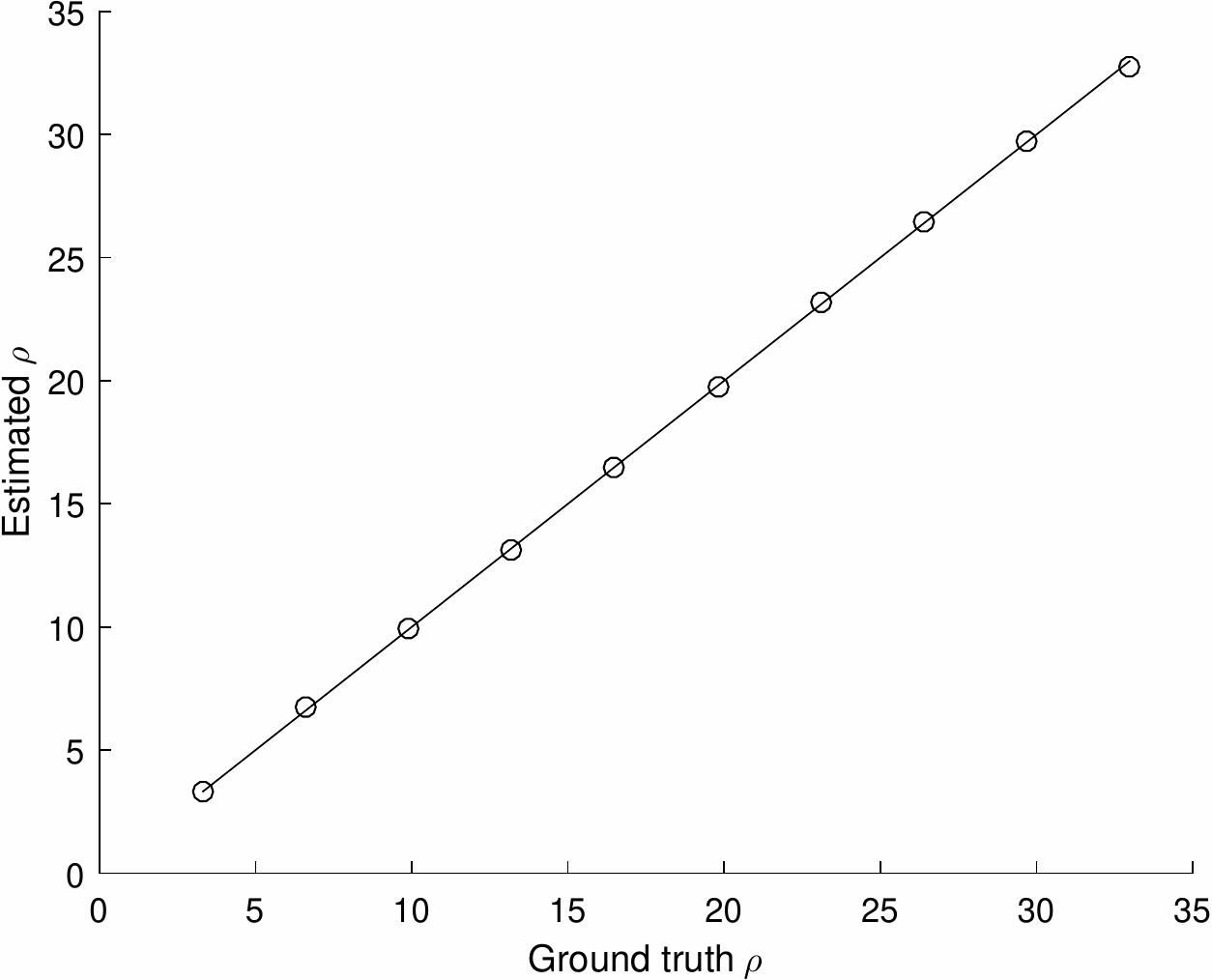}
	}
	
	\subfloat[][Deep dipole]{	\label{fig:sk_cond_est3}
		\includegraphics[width=150pt, height=150pt]{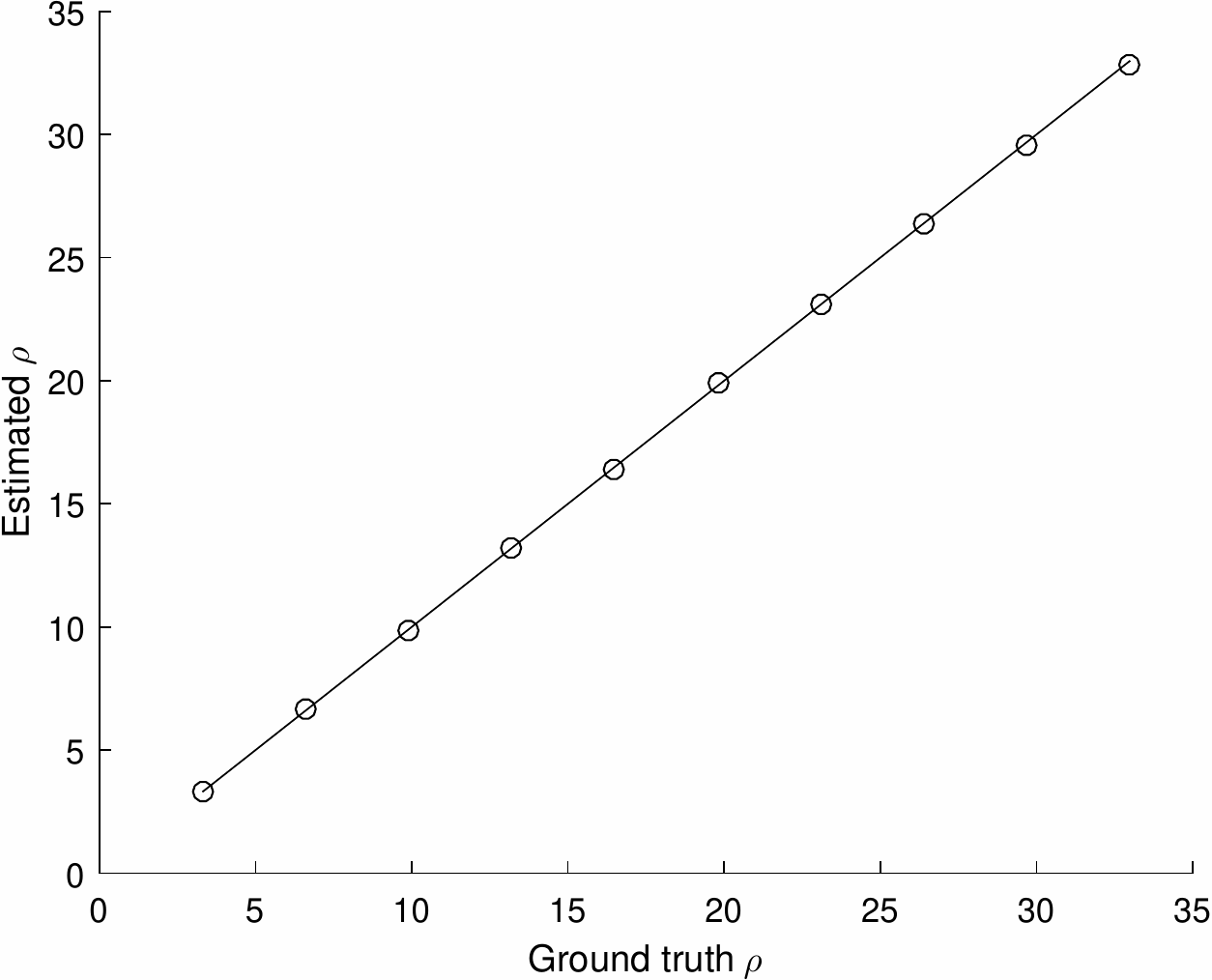}
	}
	\subfloat[][Percentage error as a function of $\rho$]{\label{fig:sk_cond_err}
		\includegraphics[width=150pt, height=150pt]{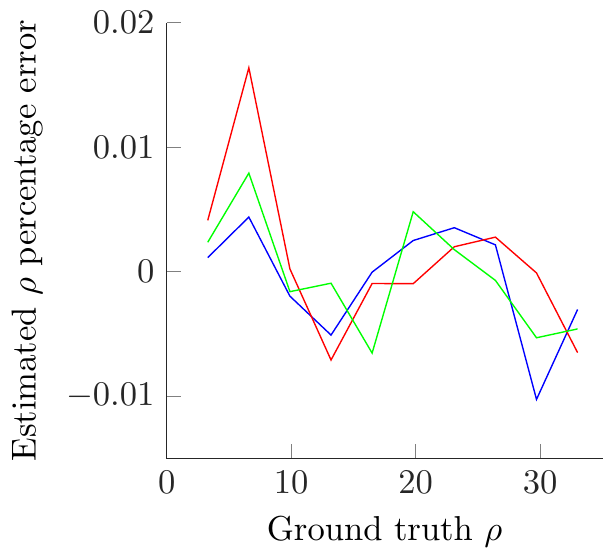}
	}
	\caption{Estimated conductivity for the multiple dipole depths experiments (SNR = 30dB).}
	\label{fig:err_30db}
\end{figure}

\begin{figure}[]
	\centering
	\subfloat[][Superficial dipole]{\label{fig:sk_cond_est1}
		\includegraphics[width=150pt, height=150pt]{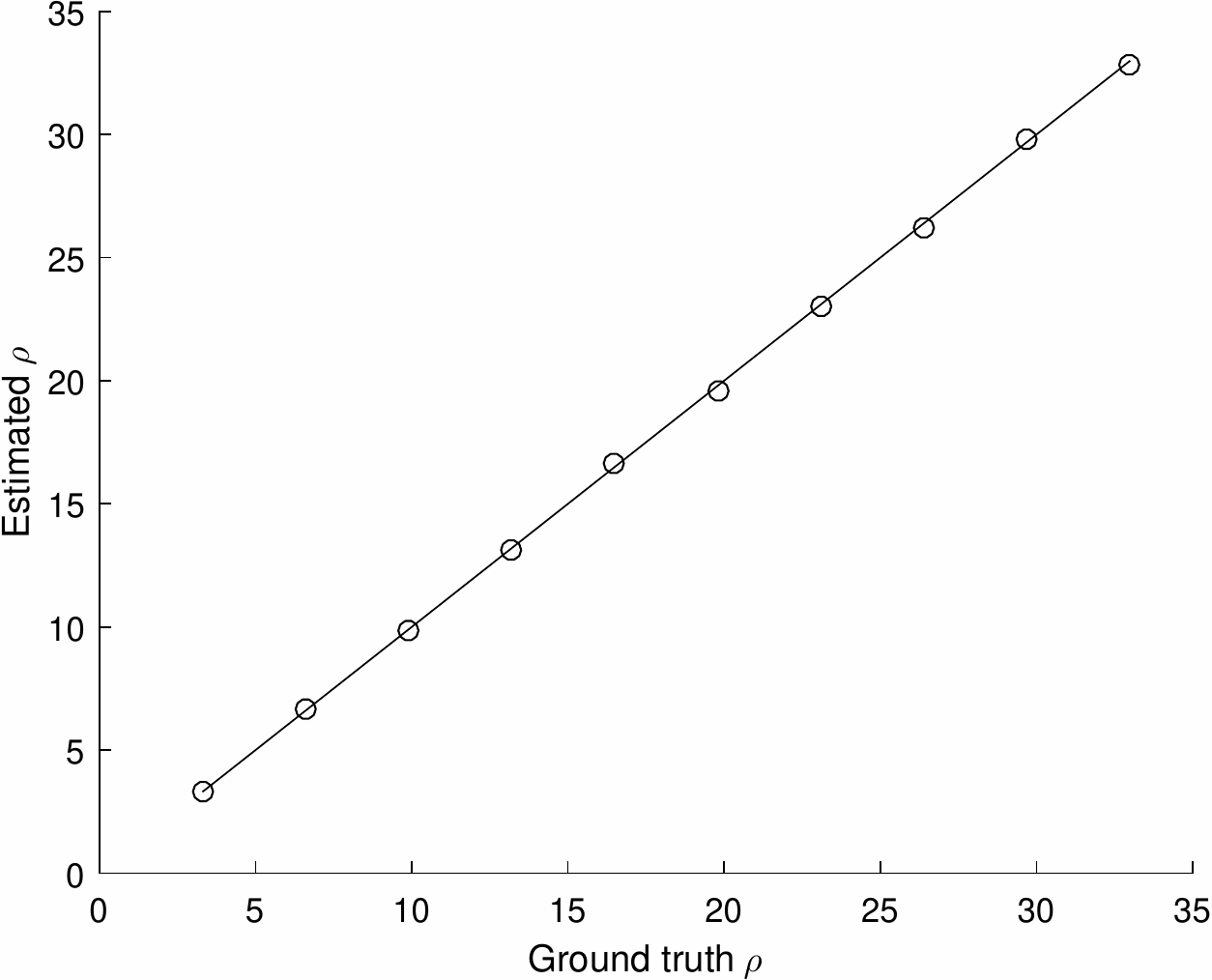}
	}
	\subfloat[][Medium dipole]{\label{fig:sk_cond_est2}	
		\includegraphics[width=150pt, height=150pt]{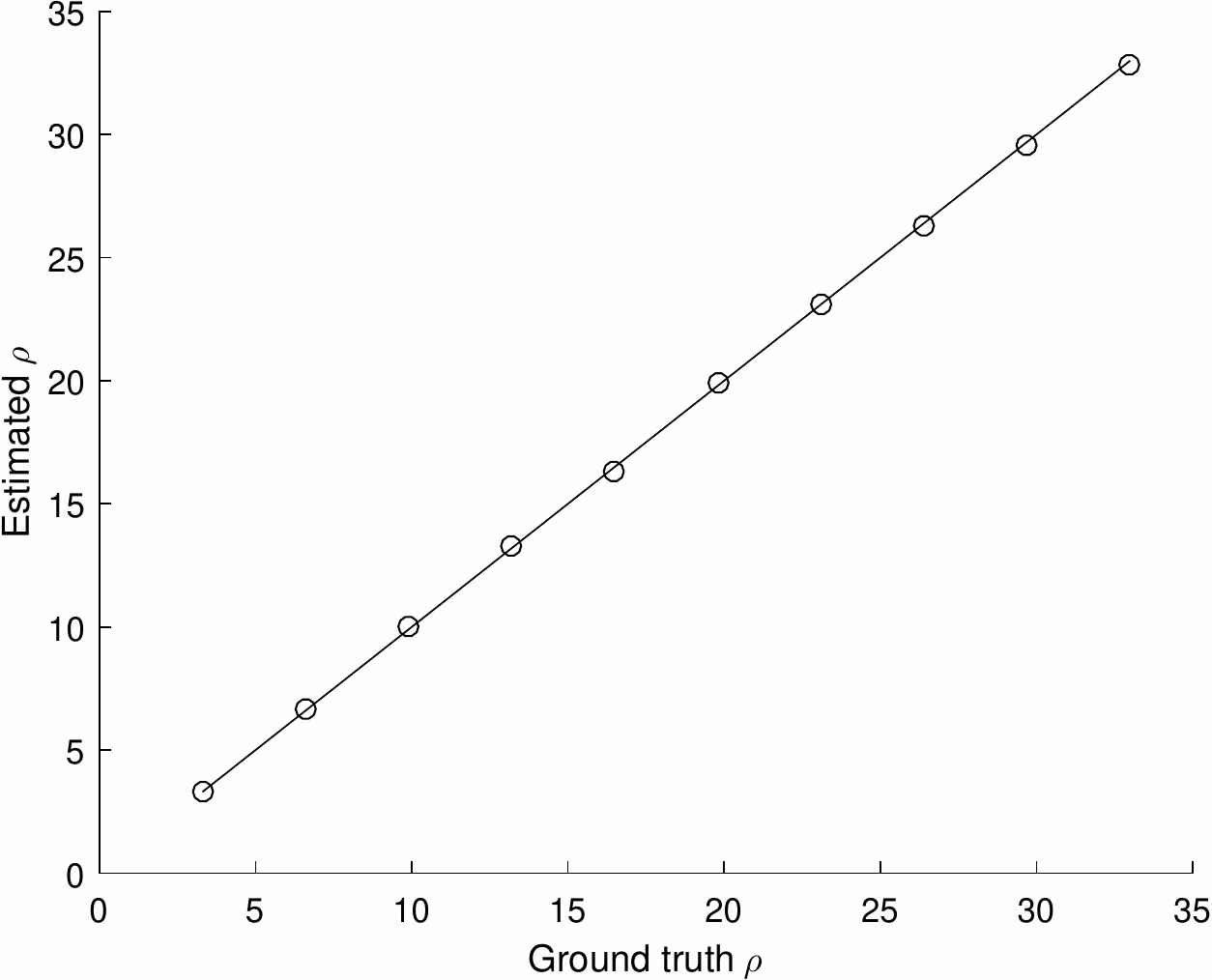}
	}
	
	\subfloat[][Deep dipole]{	\label{fig:sk_cond_est3}
		\includegraphics[width=150pt, height=150pt]{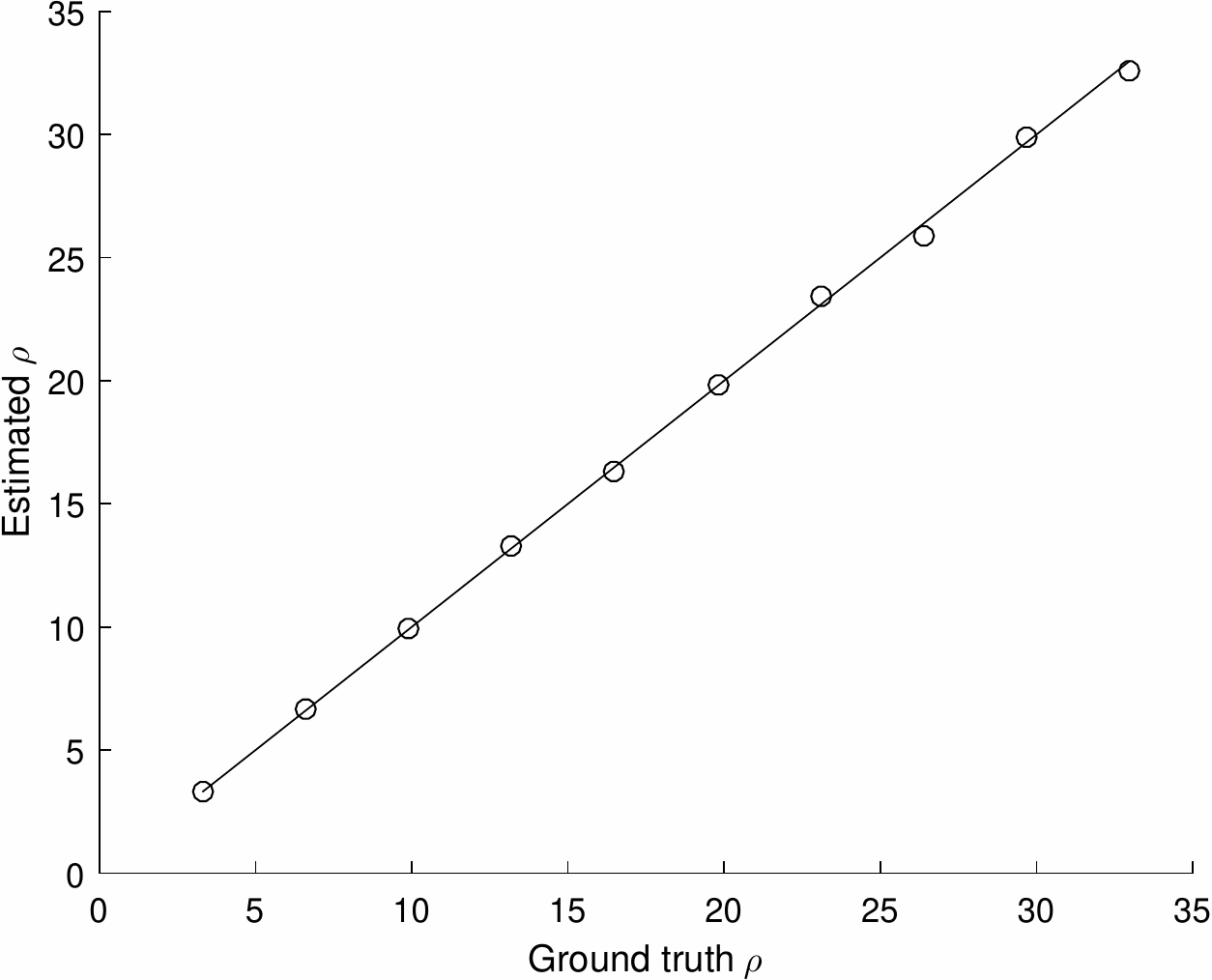}
	}
	\subfloat[][Percentage error as a function of $\rho$]{\label{fig:sk_cond_err}
		\includegraphics[width=150pt, height=150pt]{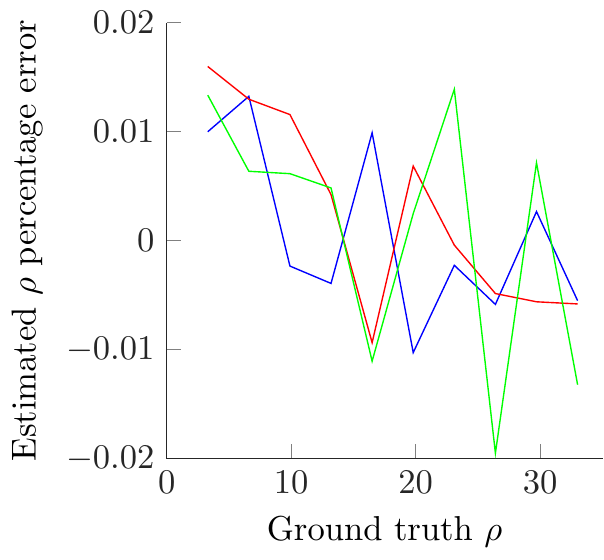}
	}
	\caption{Estimated conductivity for the multiple dipole depths experiments (SNR = 20dB).}
	\label{fig:err_20db}
\end{figure}

\begin{figure}[]
	\centering
	\subfloat[][Superficial dipole]{\label{fig:sk_cond_est1}
		\includegraphics[width=150pt, height=150pt]{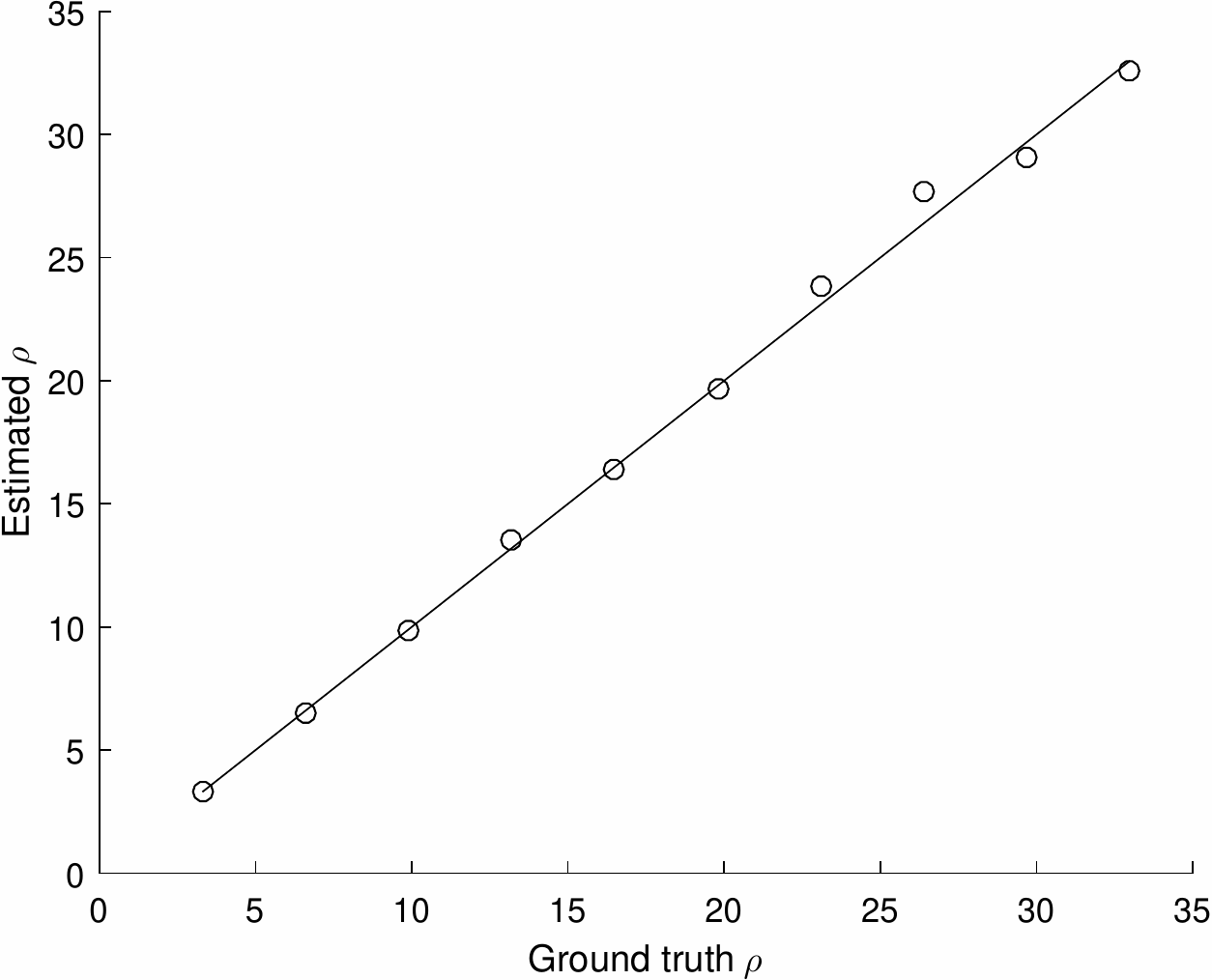}
	}
	\subfloat[][Medium dipole]{\label{fig:sk_cond_est2}	
		\includegraphics[width=150pt, height=150pt]{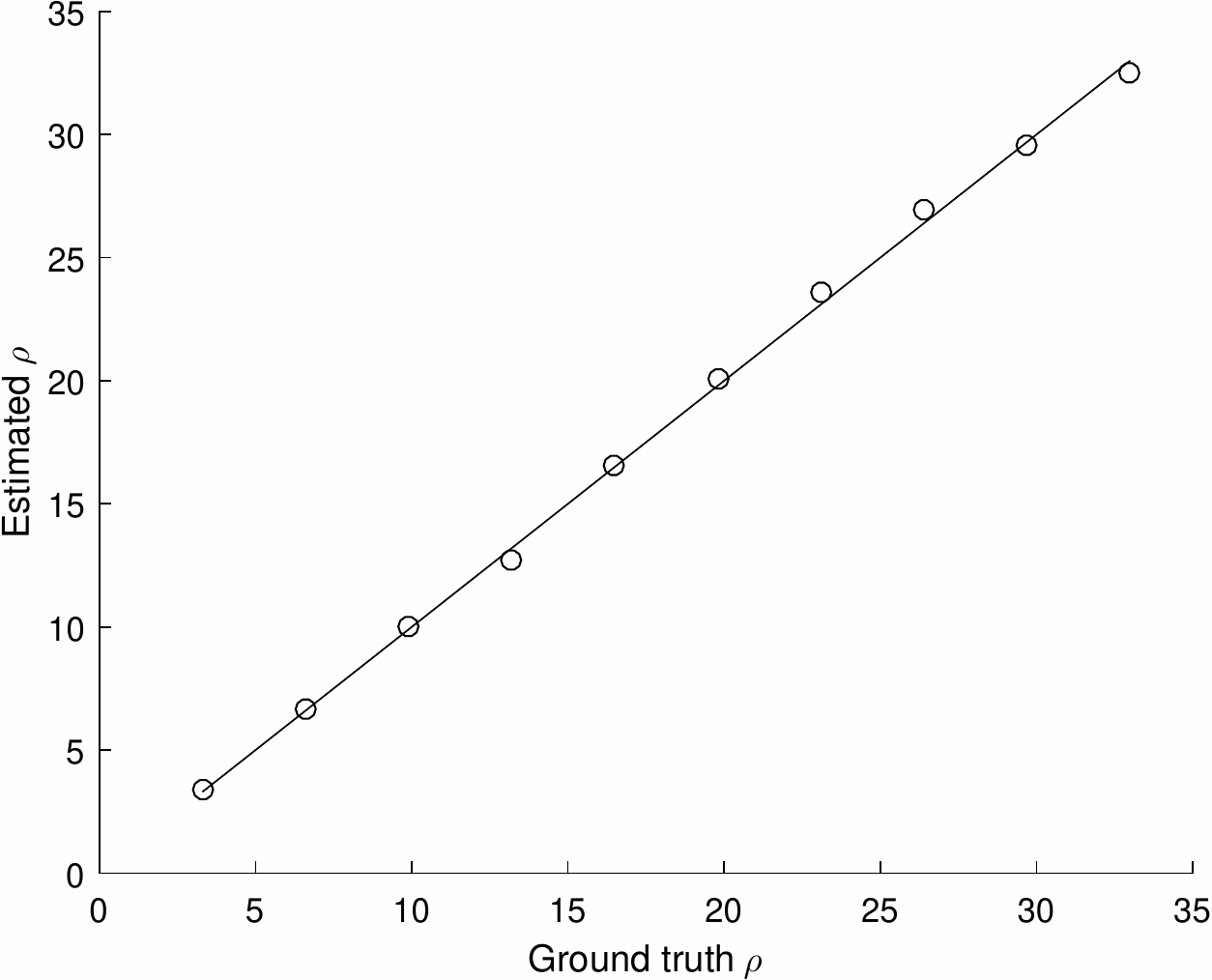}
	}
	
	\subfloat[][Deep dipole]{	\label{fig:sk_cond_est3}
		\includegraphics[width=150pt, height=150pt]{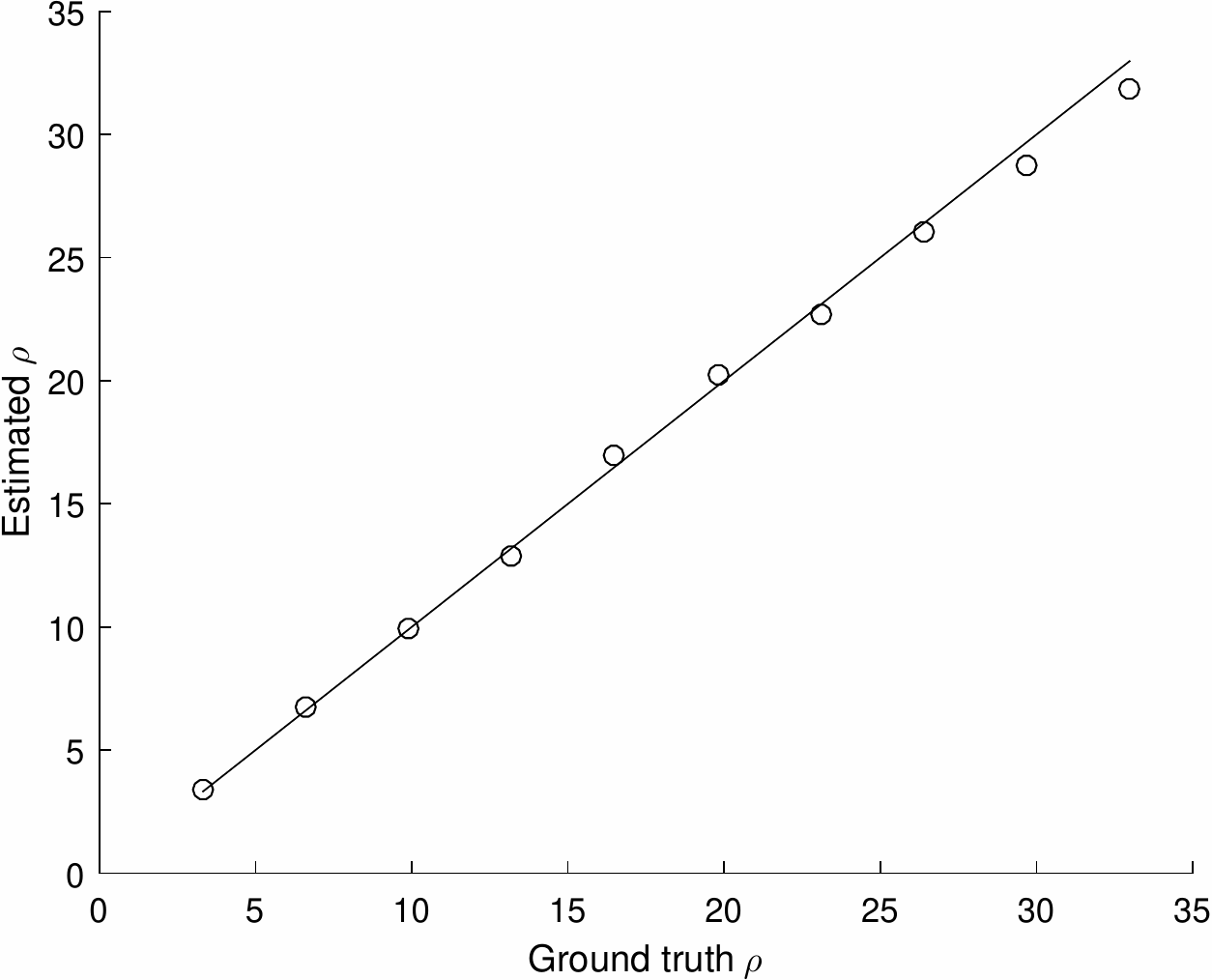}
	}
	\subfloat[][Percentage error as a function of $\rho$]{\label{fig:sk_cond_err}
		\includegraphics[width=150pt, height=150pt]{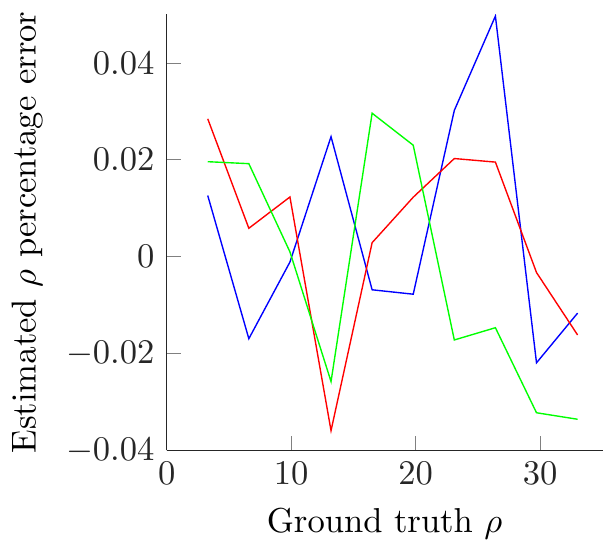}
	}
	\caption{Estimated conductivity for the multiple dipole depths experiments (SNR = 15dB).}
	\label{fig:err_15db}
\end{figure}

\begin{figure}[]
	\centering
	\subfloat[][Superficial dipole]{\label{fig:sk_cond_est1}
		\includegraphics[width=150pt, height=150pt]{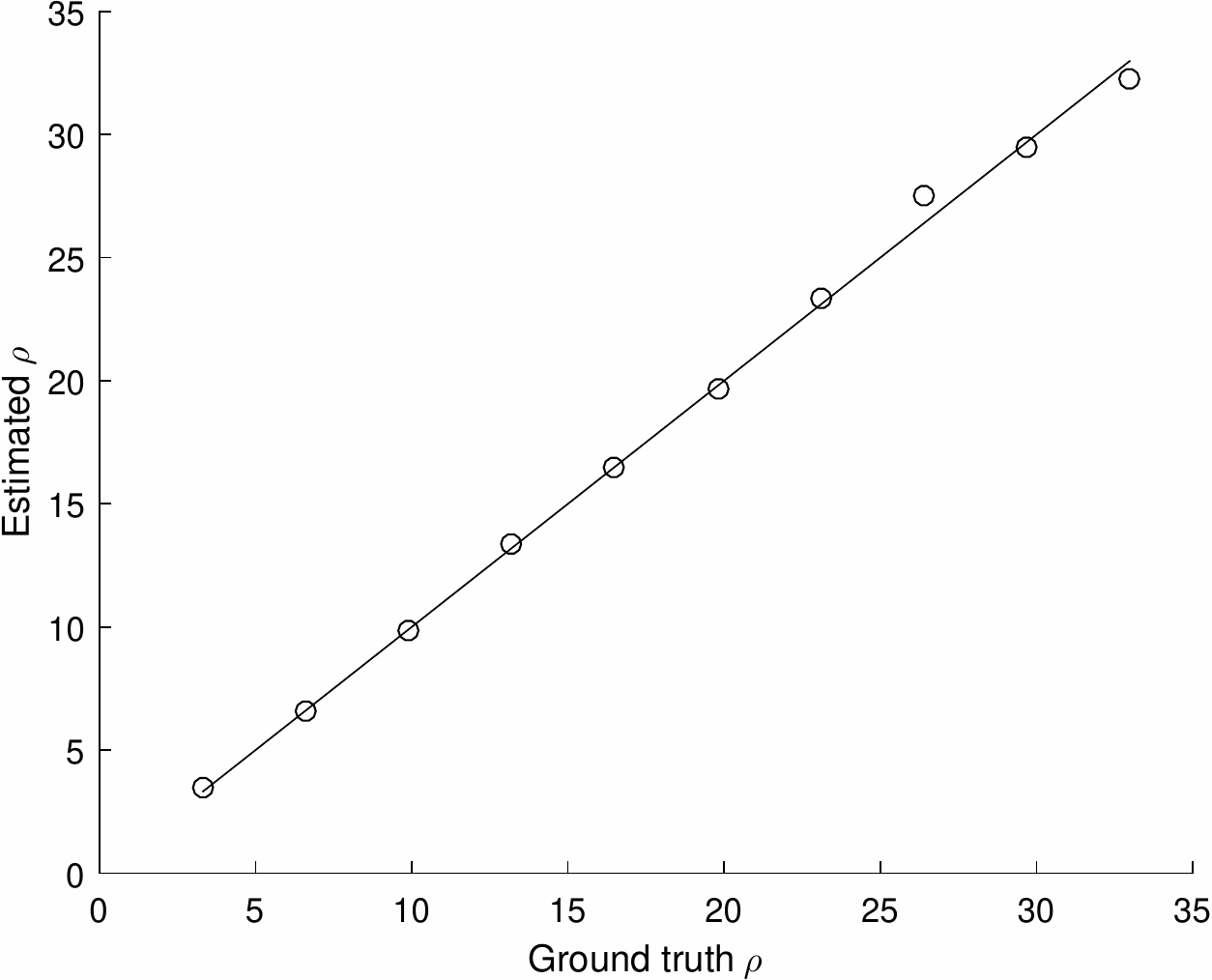}
	}
	\subfloat[][Medium dipole]{\label{fig:sk_cond_est2}	
		\includegraphics[width=150pt, height=150pt]{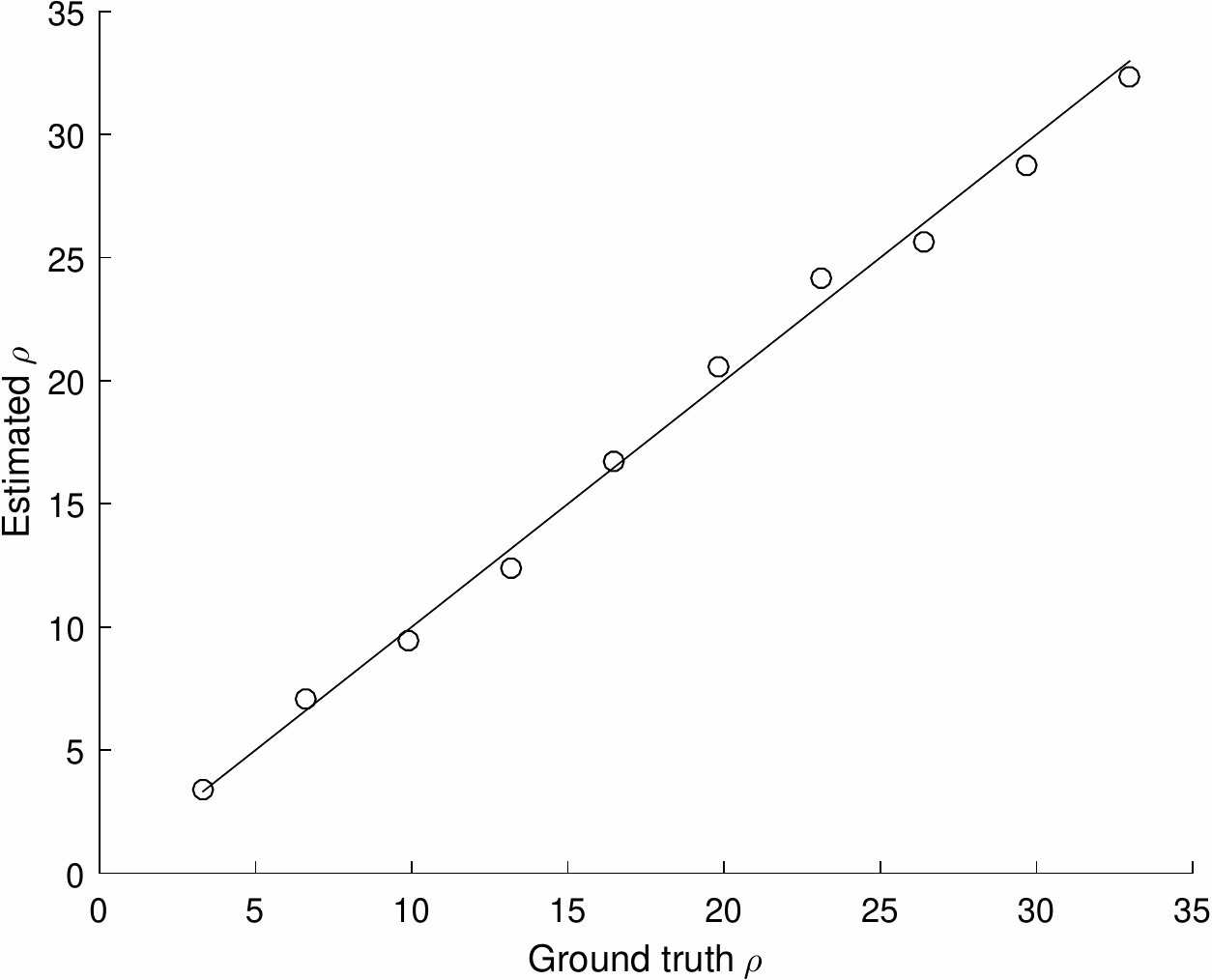}
	}
	
	\subfloat[][Deep dipole]{	\label{fig:sk_cond_est3}
		\includegraphics[width=150pt, height=150pt]{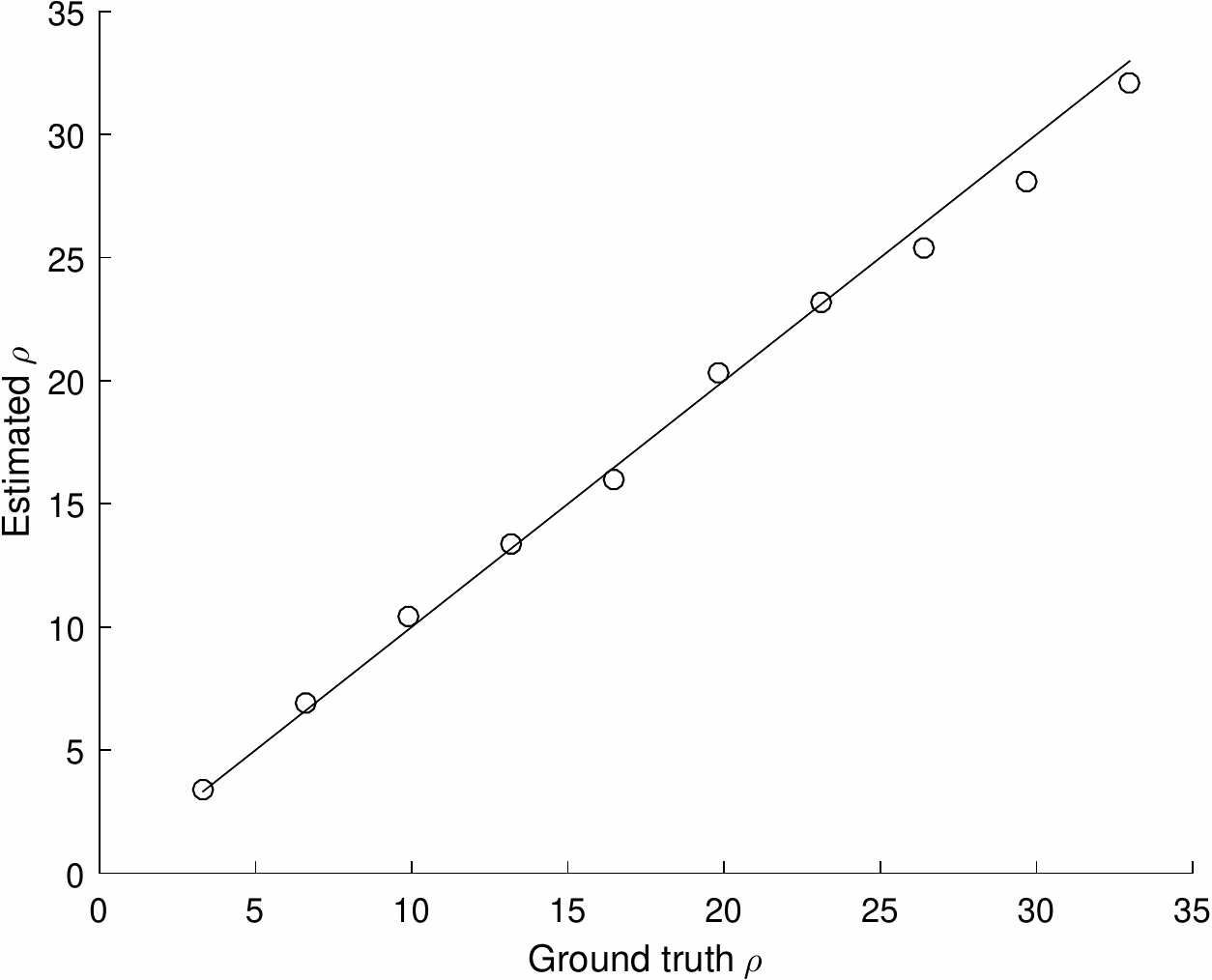}
	}
	\subfloat[][Percentage error as a function of $\rho$]{\label{fig:sk_cond_err}
		\includegraphics[width=150pt, height=150pt]{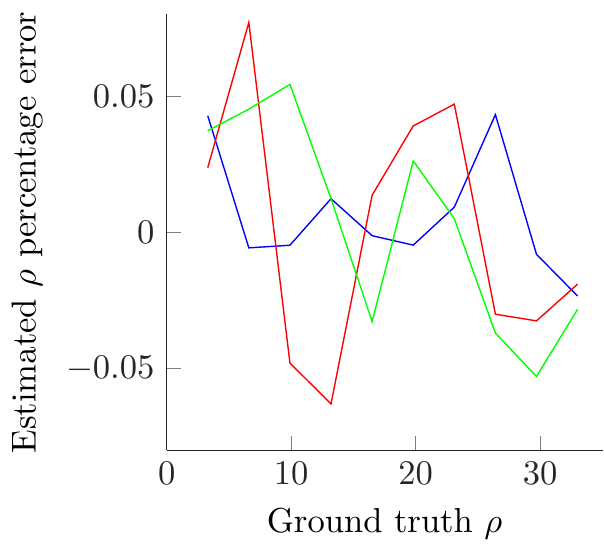}
	}
	\caption{Estimated conductivity for the multiple dipole depths experiments (SNR = 10dB).}
	\label{fig:err_10db}
\end{figure}

In all 120 experiments, the proposed method was able to correctly estimate that there was a single active dipole and which one it was (without any confusion with ones that were slightly deeper or slightly more superficial than the active one). Figs. \ref{fig:err_30db} and \ref{fig:err_20db} show the estimation errors of $\rho$ for SNR values of 30 and 20dB respectively. As we can see, the estimated values of $\rho$ are all very close to the expected ideal diagonal for all dipole depths. This can also be seen in Figs.\ref{fig:err_30db}d and \ref{fig:err_20db}d where the normalized error (expressed in percentage) is always below 2\%. In comparison, Figs. \ref{fig:err_15db} and \ref{fig:err_10db} show the results when the SNR is 15 and 10dB respectively. As expected, increasing the noise in the system makes the estimation worse, causing the normalized error to be upto 6\% for 15dB and upto 10\% for 10dB of SNR. 

These results indicate that the method is capable of estimating the depth of the active dipole and the skull conductivity jointly as long as the amount of noise present in the system is in a reasonable range.

\subsubsection{Multiple dipole activations}
The second kind of simulations considered a variable amount of active dipoles to analyze the detection capabilities of the algorithm. In each simulation, $C$ dipoles were activated with damped sinusoidal waves with frequencies varying between $5$ and $20$Hz. The activations were sampled at $200$Hz and scaled in amplitude in order to produce the same energy in the different measurements. Twenty different simulations where conducted for each value of $C=1,..., 16$, each one having a different set of active dipoles and a different uniform random value of 
conductivity in the range $\rho_{\min} < \rho_{gt} < \rho_{\max}$, resulting in a total of $320$ experiments. Noise was added to the measurements to obtain SNR = $30$dB.

Since Vallagh\'e's method cannot be applied to activations that have more than a single active dipole the performance of the proposed model was compared with two other recovery methods: (1) the correct-$\rho$ model with $\rho_{fix} = \rho_{gt}$ (to evaluate the loss of performance when $\rho$ is estimated) and (2) a default-$\rho$ model with a $\rho_{fix} = \frac {\rho_{\min} + \rho_{\max}} {2} =  18.15 \frac{mS} {m}$ (to illustrate the improvement due to the estimation of $\rho$). All models were run using 8 MCMC parallel chains. For each simulation result, the $C$ estimated active dipoles that generated the strongest measurements in $\Yb$ were considered to be the main dipoles recovered by the algorithm, while the other dipoles were used to compute the residual activity.

We define the recovery rate as the proportion of active dipoles that were recovered by a given method. Fig. \subref*{fig:rec_rate} displays the average recovery rate as a function of $C$ for the three models. The correct-$\rho$ model is able to correctly recover the dipoles perfectly up to $10$ dipoles, whereas its performance declines significantly when $C > 10$. The fact that the recovery performance decreases with more active non-zeros is well known since the operator span limits the maximum possible amount of non-zeros that can be recovered correctly \cite{candes2008restricted}. In comparison, the proposed model is able to estimate $\rho$ jointly with the brain activity up to $C = 10$ practically without any recovery loss. For $C > 10$ its performance decreases faster than the other method because of the increasing error in the estimation of $\rho$ (as shown in Fig. \subref*{fig:pcee}). The recovery rate of the default-$\rho$ model is significantly worse than the proposed method even for low values of $C$ for the same reasons mentioned for single dipoles. Since the recovery rate only considers the main detected dipoles, it is also interesting to analyze the energy contained in the residual dipoles as well. Thus, we define the proportion of residual energy as the amount of energy contained in the measurements generated by the residual activity over the total energy in the measurements. This quantity is displayed in Fig. \subref*{fig:res_energy} where we can see that both the correct-$\rho$ model and the proposed method have almost-zero energy (lower than 1\%) in the residual dipoles for low values of $C$ contrary to the default-$\rho$ model that has a larger proportion of residual energy.

Our algorithm was also compared with Gutierrez' method \cite{gutierrez2004estimating} which estimates the activity amplitude jointly with the skull conductivity but requires knowing which are the active dipoles in advance. Because of this limitation, it makes no sense to analyze the recovery rate of Gutierrez' method but we can use it to compare the estimation of $\rho$ as shown in Fig. \subref*{fig:pcee}. We can see that the estimation of both methods is comparable up to $C = 10$ despite the fact that our algorithm is able to estimate the active dipoles positions as well. For $C > 10$ the performance of our algorithm drops because it fails to recover the active dipole positions correctly.

In summary, upto $C = 10$ the quality of the reconstruction of the proposed method is very close to the method that knows the correct value of $\rho$ in advance and significantly better than the default-$\rho$ model and the error in the estimation of $\rho$ is very close to the one of Gutierrez' method (that knows the active dipole positions in advance). For $C > 10$ the performance of the proposed method drops faster than the method that knows the ground truth value of $\rho$ since the error in the estimation of $\rho$ increases considerably.


\begin{figure}[]
	\centering
	\subfloat[][Default-$\rho$ model]{
		\includegraphics{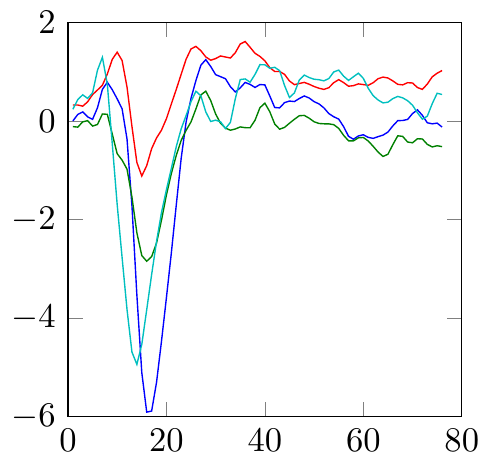}
	}
	\subfloat[][$\ell_{21}$ mixed norm]{
		\includegraphics{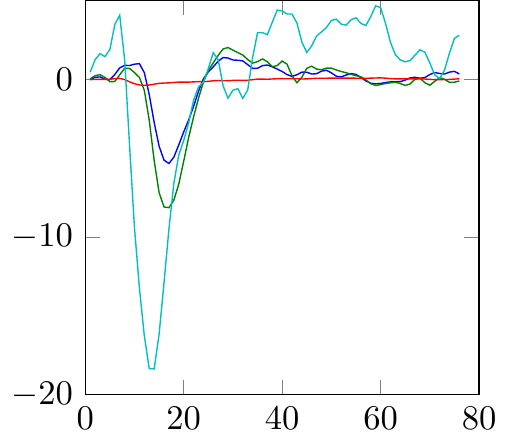}
	}
	\subfloat[][Proposed method]{
 		\includegraphics{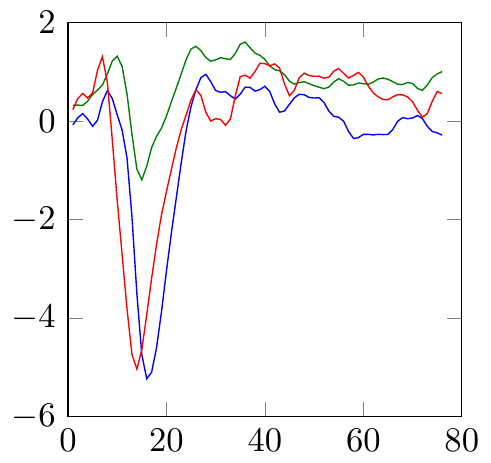}
	}
	
	\caption{Estimated waveforms for the auditory evoked responses.}
	\label{fig:real_data_waveforms_all}
\end{figure}

\begin{figure}[]
	\centering
	\subfloat[][Default-$\rho$ model]{
		\includegraphics[width=100pt, height=100pt]{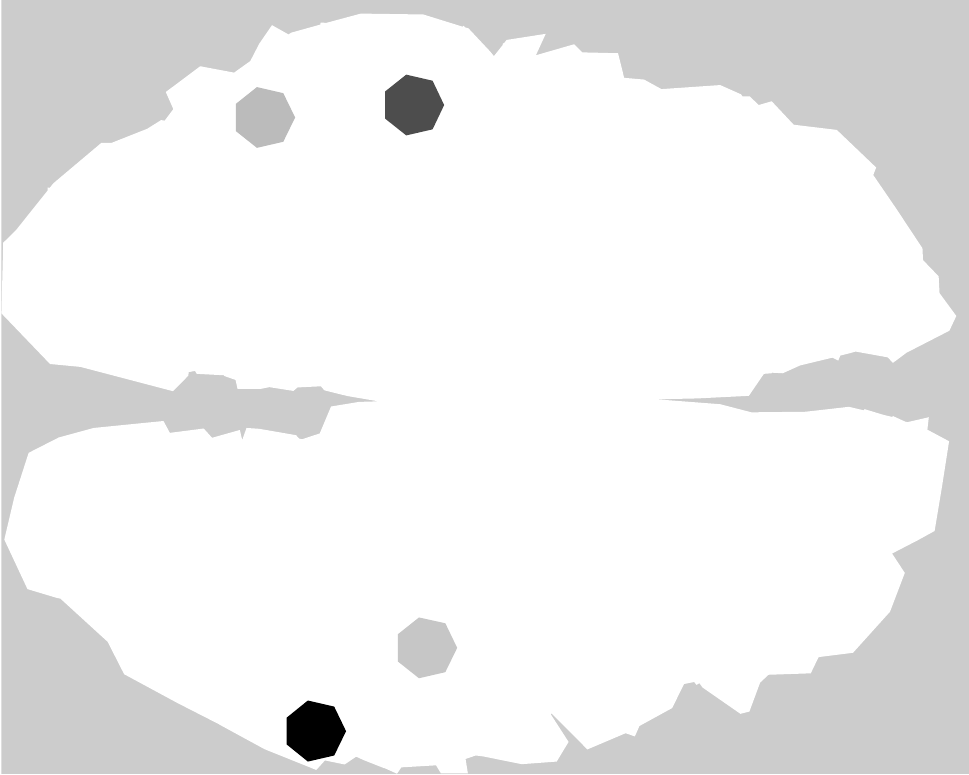}
		\includegraphics[width=100pt, height=100pt]{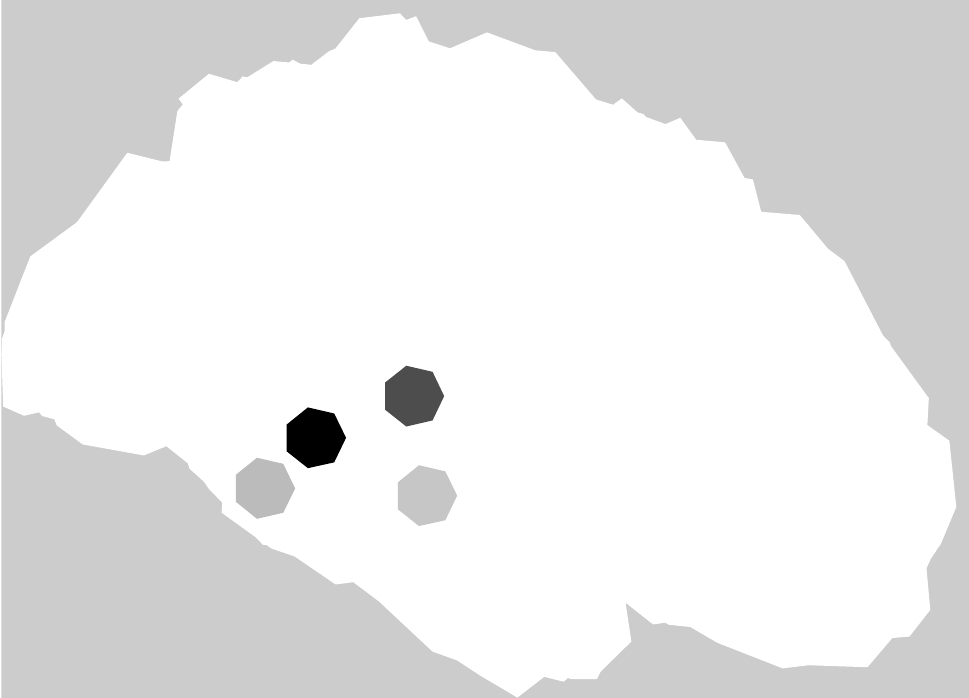}
		\includegraphics[width=100pt, height=100pt]{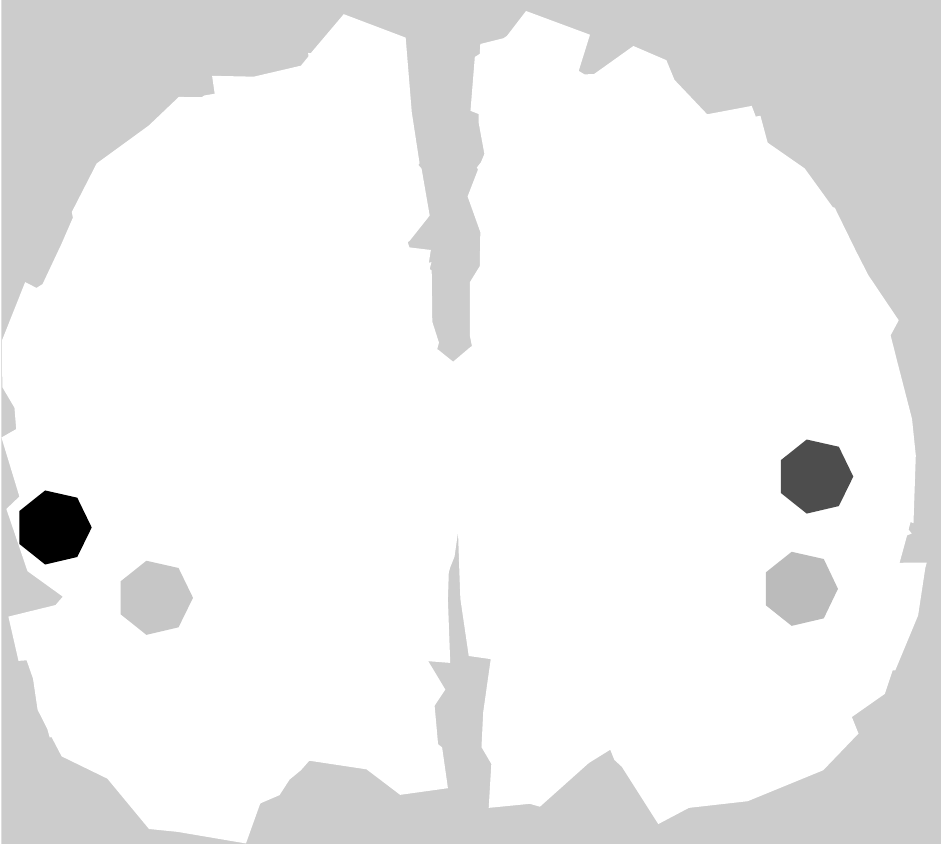}
	}

	\subfloat[][$\ell_{21}$ mixed norm]{
		\includegraphics[width=100pt, height=100pt]{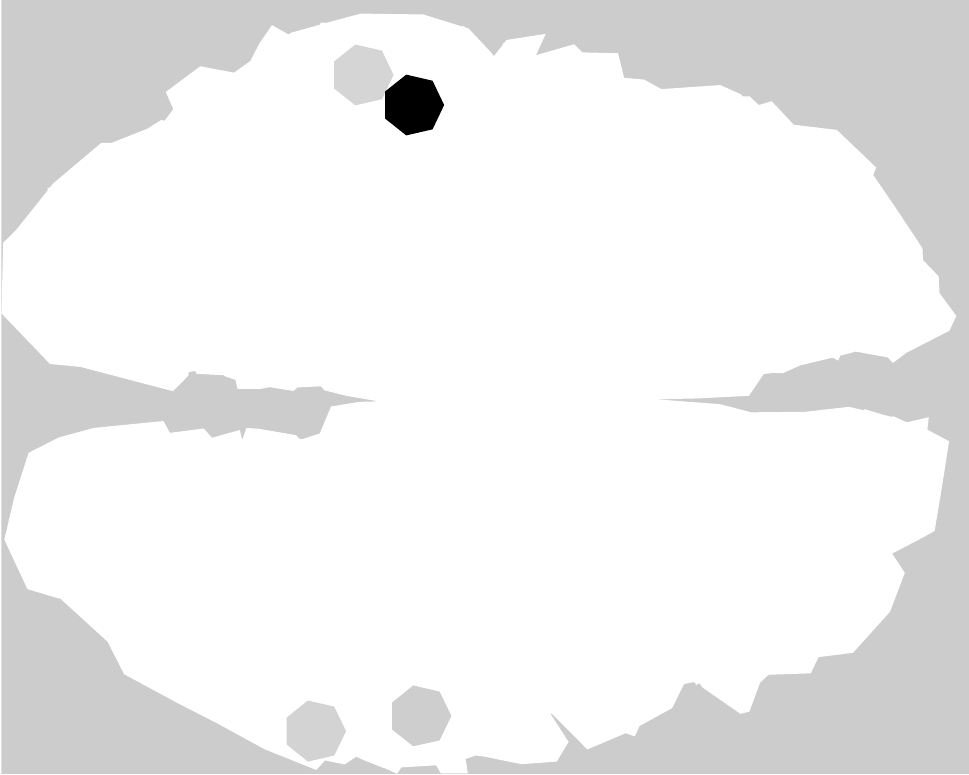}	
		\includegraphics[width=100pt, height=100pt]{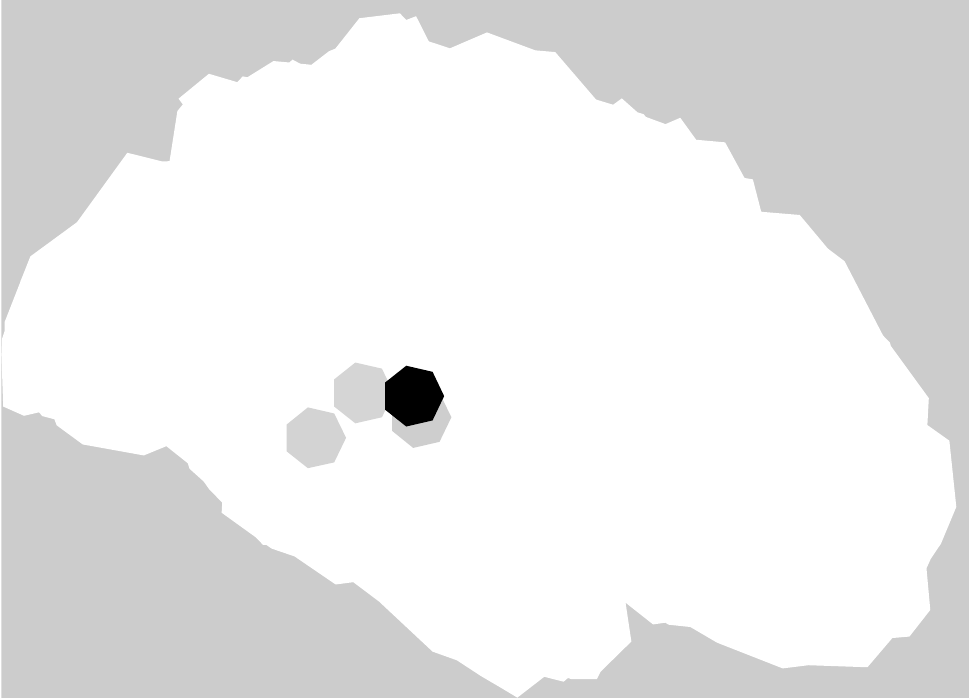}
		\includegraphics[width=100pt, height=100pt]{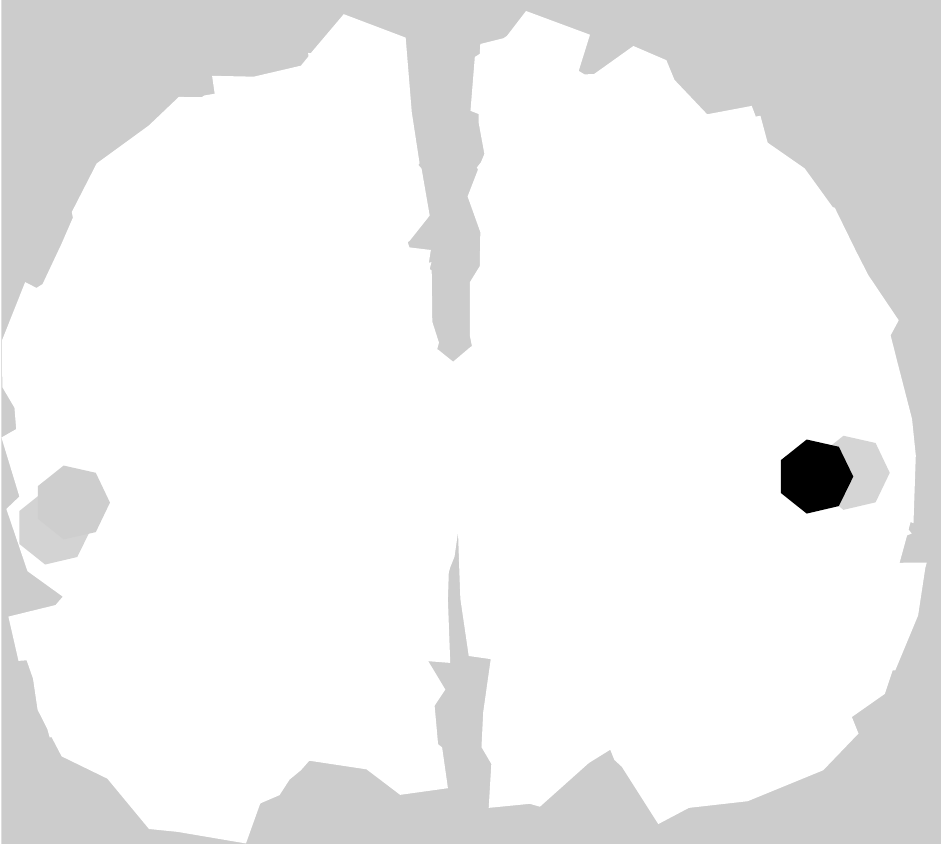}
	}

	\subfloat[][Proposed method]{
		\includegraphics[width=100pt, height=100pt]{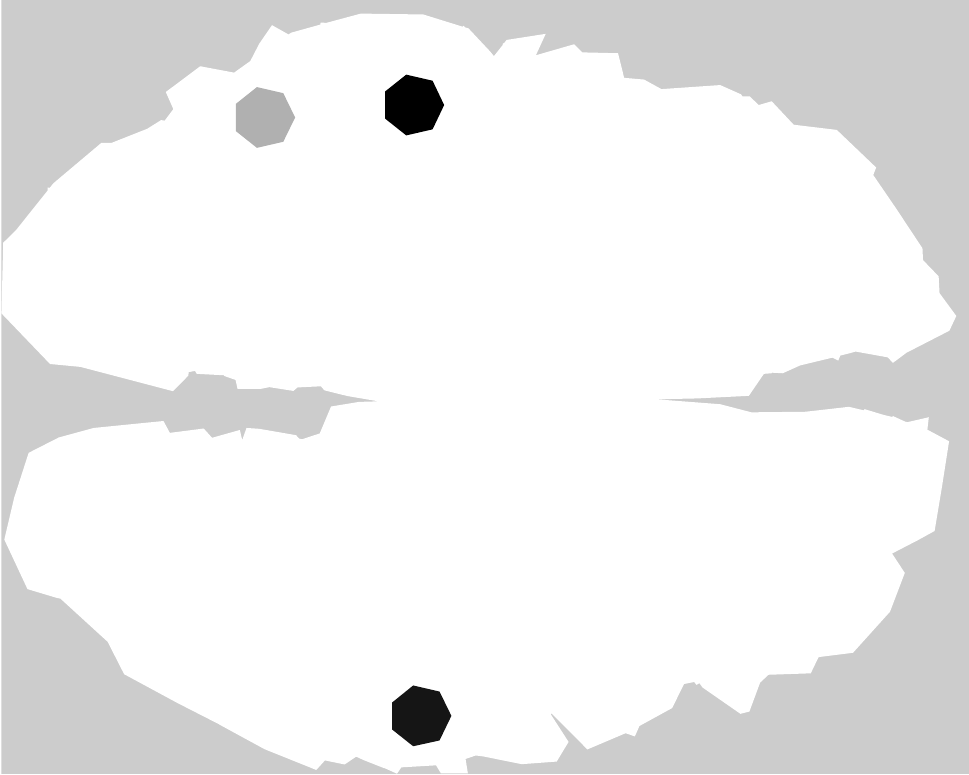}	
		\includegraphics[width=100pt, height=100pt]{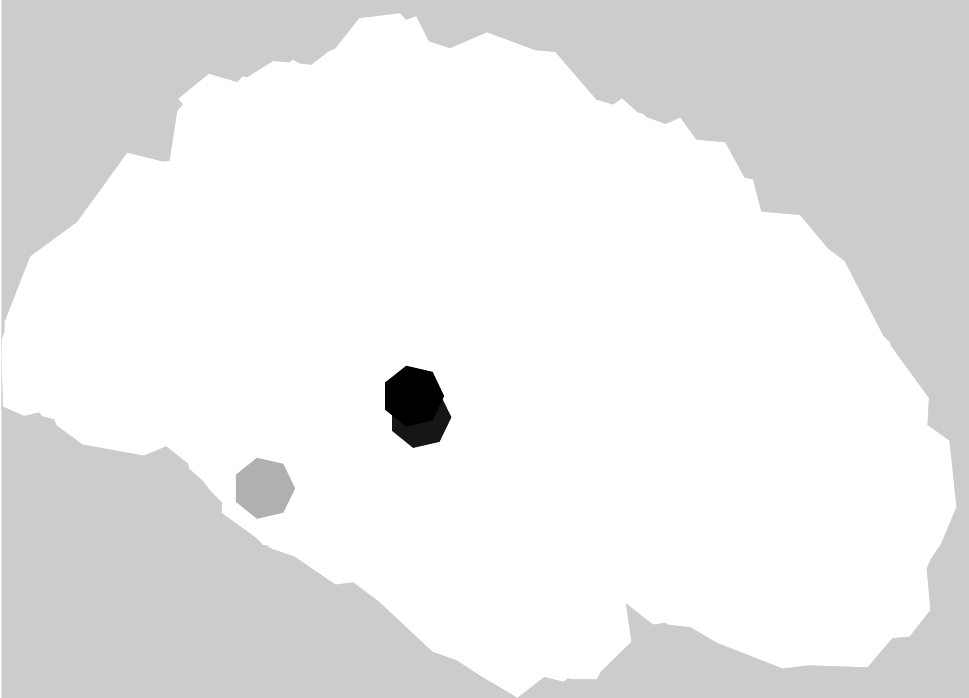}
		\includegraphics[width=100pt, height=100pt]{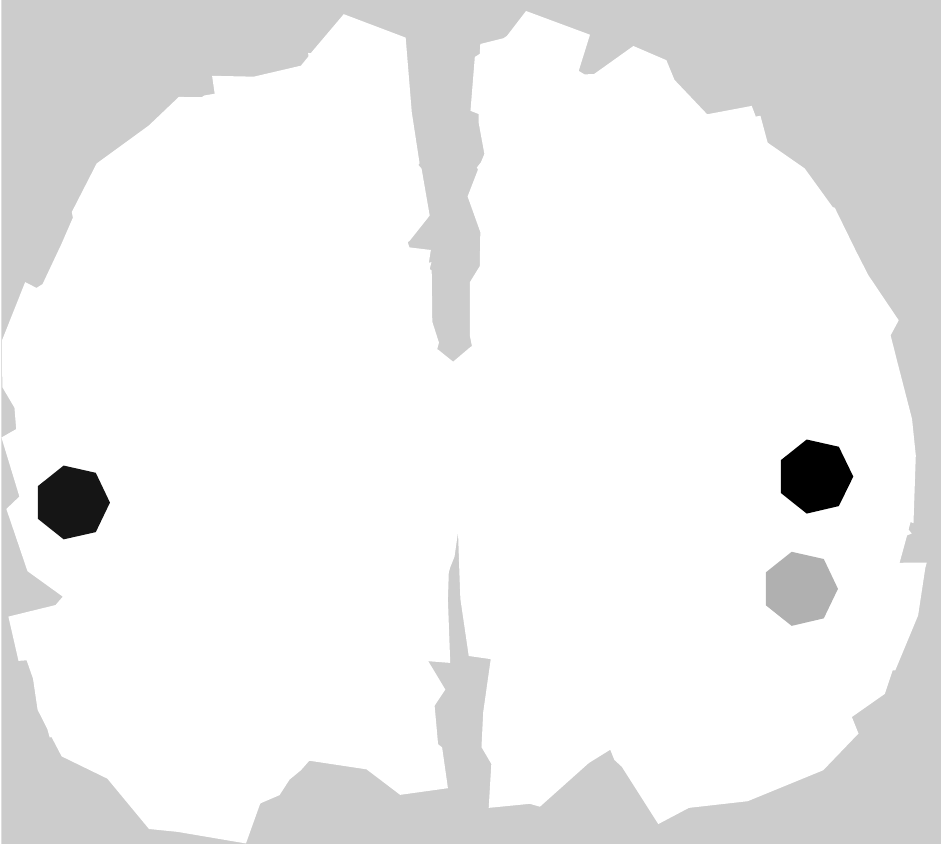}
	}
	
	\caption{Estimated activity for the auditory evoked responses.}
	\label{fig:real_data_location}
\end{figure}

\subsection{Real data}

\subsubsection{Auditory evoked responses}
The left-ear auditory pure-tone stimulus data set from the MNE software \cite{gramfort2014mne,gramfort2013meg} was first investigated. It uses a realistic BEM (Boundary element method) head model containing $1.844$ dipoles located on the cortex with orientations that are normal to the brain surface. The data was sampled with $306$ MEG sensors at $600$Hz, low-pass filtered at $40$Hz and downsampled to $150$Hz. One channel that had technical artifacts was ignored. The measurements corresponding to $200$ms of data preceding each stimulus were considered to estimate the noise covariance matrix that was used to whiten the measurements. Fifty-one epochs were averaged to calculate $\Yb$. The activity of the source dipoles was estimated jointly with the skull conductivity for the period lasting $500$ms after the stimulus. From a clinical perspective it is expected to find the brain activity primarily focused on the auditory cortices that are located close to the ears in both hemispheres of the brain.

\begin{figure}[]
	\centering
	\subfloat[][Waveform 1 (Center right dipole)]{
		\includegraphics{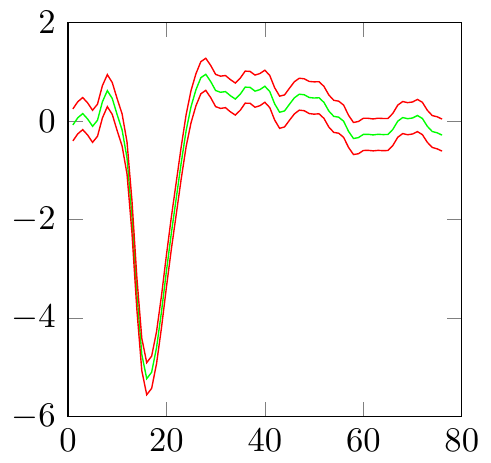}
	}
	\subfloat[][Waveform 2 (Center left dipole)]{
		\includegraphics{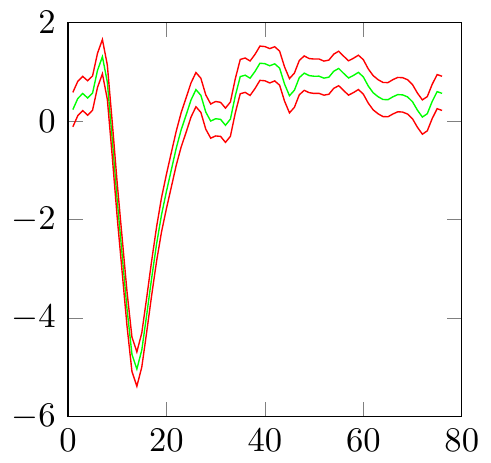}
	}
	\subfloat[][Waveform 3 (Posterior left dipole)]{
		\includegraphics{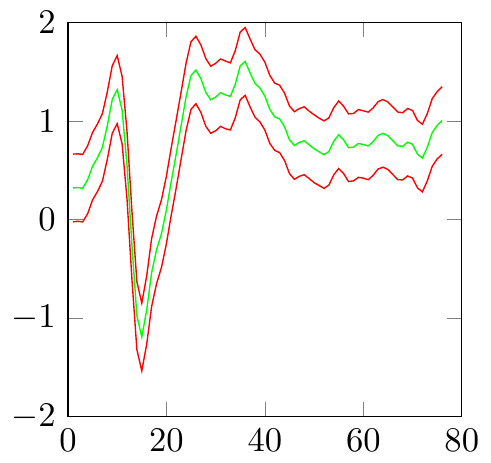}
	}
	
	\caption{Estimated waveforms mean and boundaries $\mu \pm 2 \sigma$ for the auditory evoked responses using the proposed model.}
	\label{fig:real_data_waveforms_separate}
\end{figure}

\begin{figure}[]
	\centering
	\subfloat[][Histogram of $\omega$]{
		\includegraphics{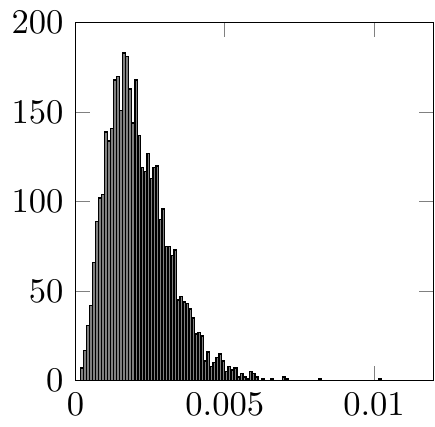}
	}
	\subfloat[][Histogram of $a$]{
		\includegraphics{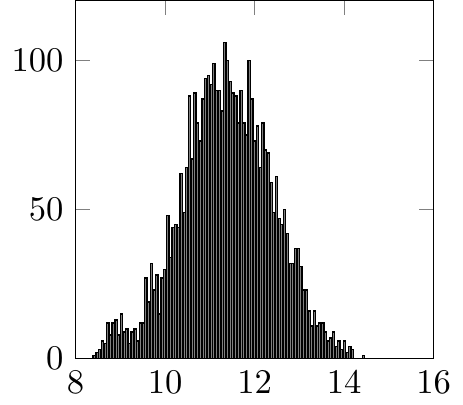}
	}	
	
	\subfloat[][Histogram of $\sigma_n^2$]{
 		\includegraphics{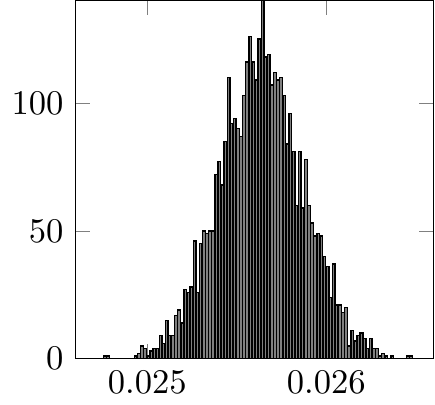}
	}
	\subfloat[][Histogram of $\rho$]{
 		\includegraphics{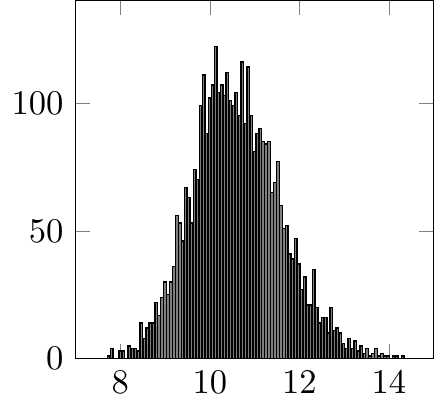}
	}

	\caption{Hyperparameter histograms for the auditory evoked responses.}
	\label{fig:real_data_histograms}
\end{figure}

The proposed method was compared with (1) a default-$\rho$ model that uses $\rho = 6 \frac {mS} {S}$, which is the value corresponding to a ratio of $50$ between the scalp and skull conductivities). It is the default value used by the MNE software, and (2) the $\ell_{21}$ mixed norm regularization introduced in \cite{gramfort2012mixed} also using an operator with $\rho = 6 \frac {mS} {S}$. 

We can see in Fig. \ref{fig:real_data_location} that out of the three models, the solution found by our method is the one that best agrees with the clinically expected solution of finding the activity focused on the auditory cortices whereas the other two spread the activity over several dipoles around the area. In addition, the MMSE estimator of the skull conductivity is $\hat{\rho} = 10.6 \frac{mS} {S}$, corresponding to a ratio of $31$ between the scalp and skull conductivities. This shows that this ratio is considerably lower than the value of $80$ typically used in earlier research \cite{geddes1967specific}, in agreement with recent studies \cite{oostendorp2000conductivity,hoekema2003measurement}.

The estimated waveforms are presented in Fig. \ref{fig:real_data_waveforms_all}: all methods estimate very similar waveforms. However, the proposed method concentrates the activity of several dipoles in the same one. The different waveforms detected by our algorithm are presented separately in Fig. \ref{fig:real_data_waveforms_separate} where the mean value for each activation and the  confidence intervals for 2 standard deviations are displayed. We can see that the two dipoles located in the auditory cortices have most of their activity concentrated in strong peaks located around $90$ ms after the stimulus, as it is clinically expected. The histograms of the sampled variables are shown in Fig. \ref{fig:real_data_histograms}.

In summary, the proposed method is able to concentrate the brain activity more strongly in the auditory cortices (where it is expected to be) than the other two and estimates a value of $\rho$ that is more compatible with the findings of recent studies than the default value used by the MNE software.

\section{Convergence Assessment}
\label{sec:convergence}
\begin{figure}[]
	\centering
	\subfloat[][PSRF of $a$]{
		\includegraphics{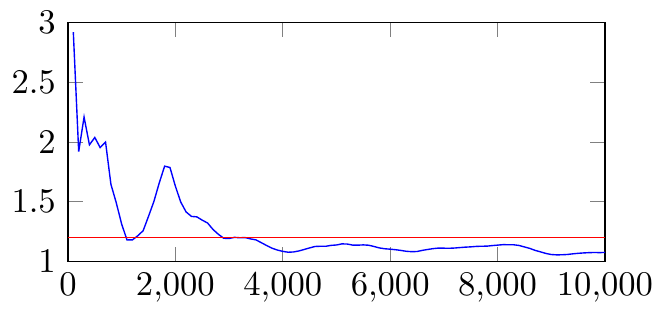}
	}
	\subfloat[][PSRF of $\omega$]{
		\includegraphics{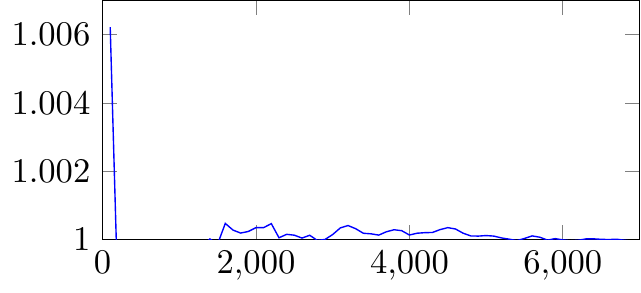}
	}
	
	\subfloat[][PSRF of $\sigma_n^2$]{
		\includegraphics{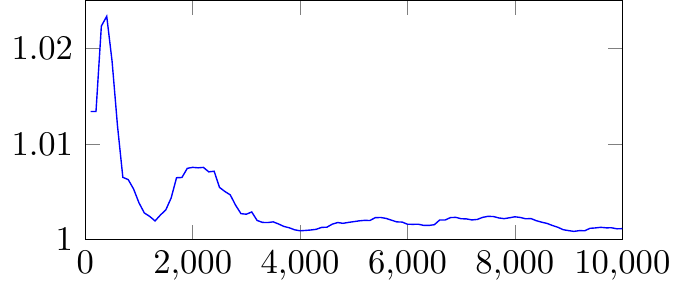}
	}
	\subfloat[][PSRF of $\rho$]{
		\includegraphics{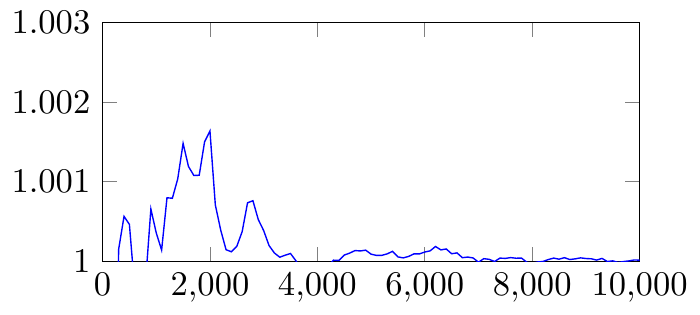}
	}
	
	\caption{PSRFs for the proposed method along the iterations for simulation $\#1$ (single dipole). Red line marks the 1.2 value that is typically considered the maximum acceptable value of PSRF that allows to conclude the simulation has converged.}
	\label{fig:single_dipole_psrfs}
\end{figure}

\begin{figure}[!]
	\centering
	\subfloat[][PSRF of $\omega$]{
		\includegraphics{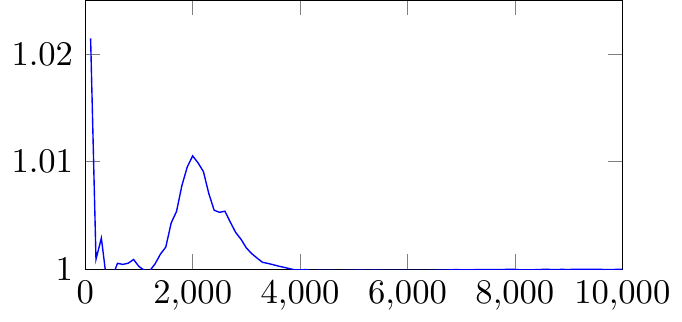}
	}
	\subfloat[][PSRF of $a$]{
		\includegraphics{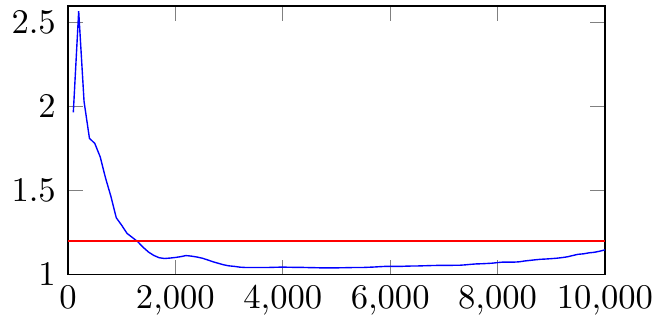}
	}
	
	\subfloat[][PSRF of $\sigma_n^2$]{
 		\includegraphics{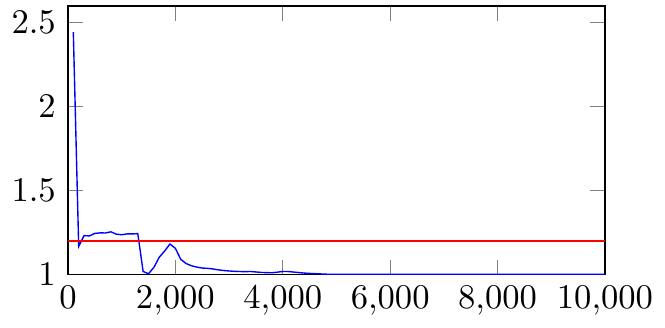}
	}
	\subfloat[][PSRF of $\rho$]{
 		\includegraphics{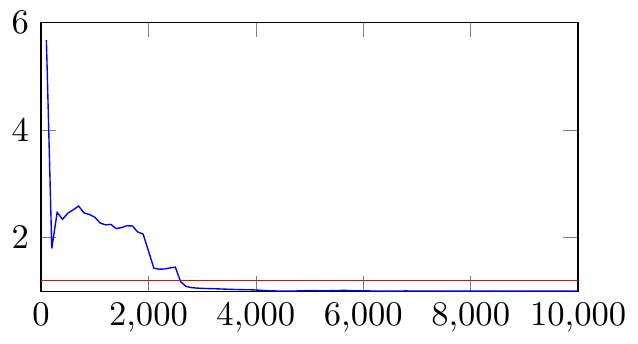}
	}		
	\caption{PSRFs of sampled variables for the auditory evoked responses along iterations. Red line marks the 1.2 value that is typically considered the maximum acceptable value of PSRF that allows to conclude the simulation has converged.}
	\label{fig:real_data_psrfs}
\end{figure}

In order to asses the convergence of the previous experiments, the Potential Scale Reduction Factors (PSRFs) \cite{brooks1998general} were considered. For single dipole simulation $\#$1, the PSRFs of the different values are displayed in Fig. \ref{fig:single_dipole_psrfs}. Fig. \ref{fig:real_data_psrfs} shows the PSRFs for the auditory evoked responses real data set. In both cases we can see that they tend to 1, showing the good numerical convergence of the proposed partially collapsed Gibbs sampler.

\section{Conclusion}
\label{sec:conclusion}
This paper introduced a novel Bayesian framework for sparse M/EEG reconstructions that is able to estimate the skull conductivity jointly with the underlying brain activity by using structured sparsity-promoting priors for the brain activity. A partially collapsed Gibbs sampler was used to generate samples from the posterior distribution that were used to estimate the model parameters and hyperparameters in a completely unsupervised framework. A polynomial approximation of the leadfield operator was used to reduce the computational cost of the sampling method used to sample the skull conductivities. The results of the proposed method were compared with the ones obtained with a fixed-conductivity model, showing that the proposed model is able to better estimate the underlying brain activity since this activity is estimated jointly with the skull conductivity. This improved performance was particularly observed when the value of skull conductivity used in the fixed-conductivity model was far away from its ground truth value or in the presence of multiple dipoles. In addition, the proposed method was compared to two different optimization techniques introduced by Vallagh\'e \textit{et al} \cite{vallaghe2007vivo} and by Gutierrez \textit{et al} \cite{gutierrez2004estimating}. Our method was shown to provide results of similar or better quality without requiring a single active dipole or knowledge of the active dipole positions in advance. Our algorithm was to an auditory evoked response real data set, showing that estimating the skull conductivity improves the quality of the reconstruction both when compared with the fixed-conductivity model and with the $\ell_{21}$ mixed norm regularization. Future work may changing the MCMC sampling to approximate message passing to reduce the computational complexity of the method.

\footnotesize
\bibliographystyle{IEEEbib}

\begin{thebibliography}{10}

\bibitem{grech2008review}
R.~Grech, T.~Cassar, J.~Muscat, K.~P. Camilleri, S.~G. Fabri, M.~Zervakis,
  P.~Xanthopoulos, V.~Sakkalis, and B.~Vanrumste,
\newblock ``{Review on solving the inverse problem in EEG source analysis},''
\newblock {\em J. Neuroeng. Rehabil.}, vol. 4, pp. 5--25, 2008.

\bibitem{hallez2007review}
H.~Hallez, B.~Vanrumste, R.~Grech, J.~Muscat, W.~De~Clercq, A.~Vergult,
  Y.~D'Asseler, K.~P. Camilleri, S.~G. Fabri, S.~Van~Huffel, et~al.,
\newblock ``{Review on solving the forward problem in EEG source analysis},''
\newblock {\em J. Neuroeng. Rehabil.}, vol. 4, pp. 46--75, 2007.

\bibitem{stok1986inverse}
C.~J. Stok,
\newblock {\em {The inverse Problem in EEG and MEG with Application to Visual
  Evoked Responses}},
\newblock Ph.D. thesis, Univ. Twente, Enschede, The Netherlands, 1986.

\bibitem{he1987electric}
B.~He, T.~Musha, Y.~Okamoto, S.~Homma, Y.~Nakajima, and T.~Sato,
\newblock ``{Electric dipole tracing in the brain by means of the boundary
  element method and its accuracy},''
\newblock {\em IEEE Trans. Biomed. Eng.}, vol. 34, no. 6, pp. 406--414, 1987.

\bibitem{johnson1995numerical}
C.~R. Johnson,
\newblock {\em Numerical methods for bioelectric field problems},
\newblock Boca Rato (FL): CRC Press, 1995.

\bibitem{marino1993finite}
F.~Marino, E.~Halgren, J.-M. Badier, M.~Gee, and V.~Nenov,
\newblock ``{A finite difference model of electric field propagation in the
  human head: Implementation and validation},''
\newblock in {\em Proc. of the 19th Annual Northeast Bioengineering
  Conference}, New Jersey, USA, 1993, IEEE, pp. 82--85.

\bibitem{wang2001influence}
Y.~Wang and J.~Gotman,
\newblock ``{The influence of electrode location errors on EEG dipole source
  localization with a realistic head model},''
\newblock {\em Clin. Neurophys.}, vol. 112, no. 9, pp. 1777--1780, 2001.

\bibitem{acar2013effects}
Z.~A. Acar and S.~Makeig,
\newblock ``{Effects of forward model errors on EEG source localization},''
\newblock {\em Brain Topogr.}, vol. 26, no. 3, pp. 378--396, 2013.

\bibitem{vallaghe2009global}
S.~Vallagh{\'e} and M.~Clerc,
\newblock ``{A global sensitivity analysis of three-and four-layer EEG
  conductivity models},''
\newblock {\em IEEE Trans. Biomed. Eng.}, vol. 56, no. 4, pp. 988--995, 2009.

\bibitem{vanrumste2000dipole}
B.~Vanrumste, G.~Van~Hoey, R.~Van~de Walle, M.~D'hav{\'e}, I.~Lemahieu, and
  P.~Boon,
\newblock ``Dipole location errors in electroencephalogram source analysis due
  to volume conductor model errors,''
\newblock {\em Med. Biol. Eng. Comput.}, vol. 38, no. 5, pp. 528--534, 2000.

\bibitem{genccer2004sensitivity}
N.~G. Gen{\c{c}}er and C.~E. Acar,
\newblock ``{Sensitivity of EEG and MEG measurements to tissue conductivity},''
\newblock {\em Phys. Med. Biol.}, vol. 49, no. 5, pp. 701, 2004.

\bibitem{geddes1967specific}
L.~Geddes and L.~Baker,
\newblock ``{The specific resistance of biological material - a compendium of
  data for the biomedical engineer and physiologist},''
\newblock {\em Med. Biol. Eng. Comput.}, vol. 5, no. 3, pp. 271--293, 1967.

\bibitem{hoekema2003measurement}
R.~Hoekema, G.~Wieneke, F.~Leijten, C.~Van~Veelen, P.~Van~Rijen, G.~Huiskamp,
  J.~Ansems, and A.~Van~Huffelen,
\newblock ``Measurement of the conductivity of skull, temporarily removed
  during epilepsy surgery,''
\newblock {\em Brain Topogr.}, vol. 16, no. 1, pp. 29--38, 2003.

\bibitem{gonccalves2003vivo}
S.~Gon{\c{c}}alves, J.~C. De~Munck, J.~Verbunt, F.~Bijma, R.~M. Heethaar,
  F.~Lopes~da Silva, et~al.,
\newblock ``{In vivo measurement of the brain and skull resistivities using an
  EIT-based method and realistic models for the head},''
\newblock {\em IEEE Trans. Biomed. Eng.}, vol. 50, no. 6, pp. 754--767, 2003.

\bibitem{lai2005estimation}
Y.~Lai, W.~Van~Drongelen, L.~Ding, K.~Hecox, V.~Towle, D.~Frim, and B.~He,
\newblock ``Estimation of in vivo human brain-to-skull conductivity ratio from
  simultaneous extra-and intra-cranial electrical potential recordings,''
\newblock {\em Clin. Neurophys.}, vol. 116, no. 2, pp. 456--465, 2005.

\bibitem{gutierrez2004estimating}
D.~Guti{\'e}rrez, A.~Nehorai, and C.~H. Muravchik,
\newblock ``{Estimating brain conductivities and dipole source signals with EEG
  arrays},''
\newblock {\em IEEE Trans. Biomed. Eng.}, vol. 51, no. 12, pp. 2113--2122,
  2004.

\bibitem{csengul2012extended}
G.~{\c{S}}eng{\"u}l and U.~Baysal,
\newblock ``{An extended Kalman filtering approach for the estimation of human
  head tissue conductivities by using EEG data: a simulation study},''
\newblock {\em Physiol. Meas.}, vol. 33, no. 4, pp. 571, 2012.

\bibitem{vallaghe2007vivo}
S.~Vallagh{\'e}, M.~Clerc, and J.-M. Badier,
\newblock ``{In vivo conductivity estimation using somatosensory evoked
  potentials and cortical constraint on the source},''
\newblock in {\em Proc. IEEE 4th Int. Symp. Biomed. Imagi. (ISBI)}, Washington
  D.C., USA, 2007, pp. 1036--1039.

\bibitem{lew2009improved}
S.~Lew, C.~H. Wolters, A.~Anwander, S.~Makeig, and R.~S. MacLeod,
\newblock ``{Improved EEG source analysis using low-resolution conductivity
  estimation in a four-compartment finite element head model},''
\newblock {\em Hum. Brain Mapp.}, vol. 30, no. 9, pp. 2862--2878, 2009.

\bibitem{lew2007low}
S.~Lew, C.~Wolters, A.~Anwander, S.~Makeig, and R.~MacLeod,
\newblock ``Low resolution conductivity estimation to improve source
  localization,''
\newblock in {\em International Congress Series}. Elsevier, 2007, vol. 1300,
  pp. 149--152.

\bibitem{costa2014l20technicalreport}
F.~Costa, H.~Batatia, T.~Oberlin, and J.-Y. Tourneret,
\newblock ``{Bayesian Structured Sparsity Priors for EEG Source Localization
  Technical Report},''
\newblock Tech. {R}ep., University of Toulouse, ENSEEIHT, 2015.

\bibitem{gramfort2012mixed}
A.~Gramfort, M.~Kowalski, and M.~H{\"a}m{\"a}l{\"a}inen,
\newblock ``{Mixed-norm estimates for the M/EEG inverse problem using
  accelerated gradient methods},''
\newblock {\em Phys. Med. Biol.}, vol. 57, no. 7, pp. 1937, 2012.

\bibitem{oostendorp2000conductivity}
T.~F. Oostendorp, J.~Delbeke, and D.~F. Stegeman,
\newblock ``The conductivity of the human skull: results of in vivo and in
  vitro measurements,''
\newblock {\em IEEE Trans. Biomed. Eng.}, vol. 47, no. 11, pp. 1487--1492,
  2000.

\bibitem{casella1999monte}
G.~Casella and C.~P. Robert,
\newblock {\em {Monte Carlo Statistical Methods}},
\newblock New York: Springer-Verlag, 1999.

\bibitem{candes2008restricted}
E.~J. Candes,
\newblock ``{The restricted isometry property and its implications for
  compressed sensing},''
\newblock {\em C. R. Acad\'emie des Sciences}, vol. 346, no. 9, pp. 589--592,
  2008.

\bibitem{gramfort2014mne}
A.~Gramfort, M.~Luessi, E.~Larson, D.~A. Engemann, D.~Strohmeier, C.~Brodbeck,
  L.~Parkkonen, and M.~S. H{\"a}m{\"a}l{\"a}inen,
\newblock ``{MNE software for processing MEG and EEG data},''
\newblock {\em NeuroImage}, vol. 86, pp. 446--460, 2014.

\bibitem{gramfort2013meg}
A.~Gramfort, M.~Luessi, E.~Larson, D.~A. Engemann, D.~Strohmeier, C.~Brodbeck,
  R.~Goj, M.~Jas, T.~Brooks, L.~Parkkonen, et~al.,
\newblock ``{MEG and EEG data analysis with MNE-Python},''
\newblock {\em Front. Neurosci.}, vol. 7, no. 267, pp. 1--13, 2013.

\bibitem{brooks1998general}
S.~P. Brooks and A.~Gelman,
\newblock ``{General methods for monitoring convergence of iterative
  simulations},''
\newblock {\em J. Comput. Graph. Statist.}, vol. 7, no. 4, pp. 434--455, 1998.

\end{thebibliography}

\end{document}